\documentclass[12pt]{iopart}
\usepackage{graphicx}
\usepackage{iopams}
\begin{document}

\title[Spatial magnetization profile in spherical nanomagnets with surface anisotropy]{Spatial magnetization profile in spherical nanomagnets with surface anisotropy: Green's function approach}
\author{M. P. Adams\footnote{michael.adams@uni.lu}, A. Michels\footnote{andreas.michels@uni.lu}}
\address{Department of Physics and Materials Science, University of Luxembourg,
162A~avenue de la Faiencerie, L-1511 Luxembourg, Grand Duchy of Luxembourg}

\author{H. Kachkachi\footnote{Corresponding author: hamid.kachkachi@univ-perp.fr}}
\address{Universit\'e de Perpignan via Domitia, Lab. PROMES CNRS UPR8521, Rambla
de la Thermodynamique, Tecnosud, 66100 Perpignan, France}

\vspace{10pt}
\begin{indented}
\item[]August 2023
\end{indented}

\begin{abstract}
We consider a single spherical nanomagnet and investigate the spatial magnetization profile $\mathbf{m}\left(\mathbf{r}\right)$ in the
continuum approach, using the Green's function formalism.
The energy of the (many-spin) nanomagnet comprises an isotropic exchange interaction, a uniaxial anisotropy in the core and N\'eel's surface anisotropy, and an external magnetic field. We derive a semi-analytical expression for the magnetization vector field $\mathbf{m}\left(\mathbf{r}\right)$ for an arbitrary position $\mathbf{r}$ within and on the boundary of the nanomagnet, as a solution of a homogeneous Helmholtz equation with inhomogeneous Neumann boundary conditions. In the absence of core anisotropy, we use the solution of this boundary problem and infer approximate analytical expressions for the components $m_{\alpha},\alpha=x,y,z$,
as a function of the radial distance $r$ and the direction solid
angle. Then, we study the effects of the nanomagnet's size and surface
anisotropy on the spatial behavior of the net magnetic moment.

In the presence of a core anisotropy, an approximate analytical solution is only available for a position $\mathbf{r}$ located on the surface,\emph{i.e.} $\mathbf{r}=R\mathbf{n}$, where $R$ is the radius of the nanomagnet
and $\mathbf{n}$ the verse of the normal to the surface. This solution
yields the maximum spin deviation as a result of the competition between
the uniaxial core anisotropy and N\'eel's surface anisotropy. Along
with these (semi-)analytical calculations, we have solved a system
of coupled Landau-Lifshitz equations written for the atomic spins,
and compared the results with the Green's function approach.

For a plausible comparison with experiments, \emph{e.g.} using the technique
of small-angle magnetic neutron scattering, we have averaged over
the direction solid angle and derived the spatial profile in terms
of the distance $r$. We believe that the predictions of the present
study could help to characterize and understand the effects of size
and surface anisotropy on the magnetization configurations in nanomagnet
assemblies such as arrays of well-spaced platelets.
\end{abstract}

\section{Introduction}
Why has studying nanoscaled systems suddenly become so important?
How can a material that is so small be so influential as to trigger
a tremendous research activity worldwide? To understand the power
of nanomaterials, we need to understand how materials work at the
very small scale. When the dimensions of the system are reduced down
to the nanoscale, the first big change is the larger proportion of
atoms at its surface, and this leads to a dramatic change in the chemical
and physical properties. For example, bulk gold is an inert metal
that does not react much and thus does not rust, whereas at the nanoscale,
it works as a catalyst. A nanoparticle of gold about 90nm in size
absorbs red and yellow light from the color spectrum, making the nanoparticle
appear blue or green \cite{bohhuf83vch}. However, the same nanoparticle
that is only 30nm in size, absorbs blue and green light, and so appears
red. Today, we know that this is so because the optical properties
of metal nanoparticles are dominated by the surface plasmon resonances
for both absorption and scattering of light \cite{Jinetal_nat03}.
At the bottom end of the nanoscale, quantum phenomena start to emerge
through a transition from the electronic-band structure to discrete
energy levels \cite{Zhuetal_jacs08}. This fundamentally alters the
material properties and leads to many new phenomena \cite{Jinetal_cr16},
such as the energy gap and excitonic absorption of light, unique catalytic
activity, and single-electron magnetism. In addition to the electronic
structure alteration, the crystal structure also starts to exhibit
significant changes with respect to the bulk material.

The reduction of the size and the entailed changes in the electronic
and crystal structure fundamentally alter the magnetic properties
as well. As the size reduces to the nanometer, the magnetic system,
nowadays called \emph{a nanomagnet}, exhibits a new phenomenon known
as \emph{superparamagnetism} \cite{nee53cras,beanlevi59jap} with
a change in the relevant temperature by an order of magnitude and
in the relaxation time by several orders of magnitude. Furthermore,
the large surface contribution in nanoscaled magnetic systems leads
to inhomogeneous atomic-spin configurations {[}see Refs. \cite{kacgar05springer,SRAJER20061,SchmoolKachkachi_acapress2016,koumpouras2017phd,IglesiasKachkachi_spring2021}
and references therein{]} and thereby to a drastically different behavior
in response to external stimuli. For example, new modes of magnetization
switching play a crucial role in many new spintronic applications
\cite{HirohataEtal_jmmm20}. Likewise, these surface-induced effects
strongly affect the relaxation processes owing to a more complex potential
energy and new excitation modes \cite{dejardinetal08jpd,Vernay_etal_acsucept_PRB2014}.

Therefore, in order to understand and master the new magnetic properties of nano-elements, in view of efficient practical applications, it
is essential to probe the novel features induced by their surfaces and interfaces. In particular, it is crucial to investigate the atomic-spin
configurations within the nanomagnets and characterize the spatial profile of their magnetization. In this work, we present a study of
the latter within spherical nanomagnets using the technique of Green's functions (GF). This is part of a broader work that makes use of complementary techniques with the aim to better characterize surface-induced spin noncollineartities and their effects on the equilibrium and dynamic behaviors of assemblies of nanomagnets.
In this context, the present work provides a general formalism that renders fairly precise and useful analytical expressions of the spatial magnetization profile, clearly presenting all the involved mathematical steps and approximations. As a byproduct, it confirms and elaborates on the conclusions reached by previous works regarding the effects of surface anisotropy in nanomagnets.

When averaged over the solid angle, the spatial magnetization profile may be compared with experiments, such
as magnetic small-angle neutron scattering (SANS)\cite{MuhlbauerEtAL_RevModPhys.91.015004}
in assemblies of nanomagnets. Indeed, the present formalism constitutes
a basis for computing the magnetic SANS cross-section of nanomagnets
as a function of their various physical parameters. Accordingly, in
a recent study \cite{AdamsEtal_jac22_ana,AdamsEtal_jac22_num}, we
investigated the signature of surface-induced spin misalignments in
the SANS cross section upon varying the applied magnetic field and
the nanomagnet energy parameters.

Plan of the article: After an introduction,
in Section \ref{sec:Many-spin-nanomagnet}, we present our model for
a nanomagnet, viewed as a crystallite of $\mathcal{N}$ atomic magnetic
moments and then describe the continuum approach for studying the
spatial profile of its magnetization. In Section \ref{sec:Magnetization-profile:GF},
we use the Green's function technique to obtain the spin deviation
vector in terms of surface anisotropy, in addition to other parameters,
both with and without the anisotropy in the core. We plot the magnetization
profile in the radial direction and study its behavior as we vary
the size of the nanomagnet and the anisotropy constants. The results
from the Green's function formalism are also compared to the numerical
calculations based on the solution of the system of coupled Landau-Lifshitz
equations. Finally, in Section \ref{sec:Summary-Conclusions}, we
summarize the main results of this work and discuss the possibility
of experimentally investigating the surface-induced spin-misalignments,
\emph{e.g.} by SANS technique. The paper ends with an Appendix.

\section{\label{sec:Many-spin-nanomagnet}Many-spin nanomagnet}

\subsection{Discrete lattice}

A many-spin nanomagnet (NM) is viewed as a crystallite of $\mathcal{\ensuremath{N}}$
atomic magnetic moments $\mathbf{\mu}_{i}=\mu_{a}\mathbf{m}_{i}$
($\left\Vert \mathbf{m}_{i}\right\Vert =1$) with $\mu_{a}=M_{s}v_{0}$,
where $M_{s}$ is the saturation magnetization and $v_{0}$ is the
volume of the unit cell of the underlying lattice. The magnetic state
of the NM may be investigated with the help of the atomistic approach
based on the anisotropic (classical) Dirac-Heisenberg Hamiltonian
\cite{dimwys94prb,kodber99prb,kacgar01physa300,kacgar01epjb,igllab01prb,kacdim02prb,kacgar05springer,kazantsevaetal08prb,Evans_2014}
\begin{eqnarray}
\mathcal{H} & =-\frac{1}{2}\sum\limits _{\left\langle i,j\right\rangle}J_{ij}\,\mathbf{m}_{i}\cdot\mathbf{m}_{j}-\mu_{a}\mathbf{H}_{\rm{ext}}
\cdot\sum_{i=1}^{\mathcal{N}}\mathbf{m}_{i} + \sum_{i=1}^{\mathcal{N}}\mathcal{H}_{{\rm an},i} \label{eq:Ham-MSP}\\
& \equiv\mathcal{H}_{{\rm exc}} + \mathcal{H}_{{\rm Z}} + \mathcal{H}_{\rm an},\nonumber
\end{eqnarray}
where $\mathcal{H}_{{\rm exc}}$ is the (nearest-neighbor) exchange
energy, $\mathcal{H}_{{\rm Z}}$ the Zeeman contribution and $\mathcal{H}_{{\rm an}}\equiv\sum_{i=1}^{\mathcal{N}}\mathcal{H}_{{\rm an},i}$
the anisotropy energy with $\mathcal{H}_{{\rm an},i}=-K_{i}\,\mathcal{A}\left(\mathbf{m}_{i}\right)$,
being the anisotropy contribution of each spin on site $i$; $\mathcal{A}\left(\mathbf{m}_{i}\right)$
is the anisotropy function that depends on the locus of the atomic
spin $\mathbf{s}_{i}$. So, for core spins, the anisotropy may be
uniaxial and/or cubic, while for surface spins there are a few models
for on-site anisotropy that is very often taken as uniaxial with either
a transverse or parallel easy axis. There is also the more plausible
model proposed by N\'eel \cite{nee54jpr} for which $\mathcal{A}\left(\mathbf{m}_{i}\right)=\frac{1}{2}\sum\limits _{j=1}^{z_{i}}\left(\mathbf{m}_{i}\cdot\mathbf{u}_{ij}\right)^{2}$,
where $z_{i}$ is the coordination number at site $i$ and $\mathbf{u}_{ij}$
a unit vector connecting the nearest neighbors $i,j$. The constant
$K_{i}>0$ is usually denoted by $K_{{\rm c}}$ if the site $i$ is
in the core and by $K_{{\rm s}}$ if it is on the boundary.

Therefore, in the sequel we will refer to the N\'eel Surface Anisotropy
(NSA) model and this means that we consider a uniaxial anisotropy
in the core with easy axis (whose verse is the unit vector $\mathbf{e}_{A}$) and N\'eel's on-site
anisotropy for spins on the surface. More precisely, in the NSA model
we adopt the following anisotropy energy

\begin{equation}
\mathcal{H}_{{\rm an},i}=\left\{
\begin{array}{lc}
-K_{c}\left(\mathbf{m}_{i}\cdot\mathbf{e}_{A}\right)^{2}, & i\in{\rm core}\\
\\
+\frac{1}{2}K_{s}{\displaystyle \sum_{j\in{\rm n.n.}}}\left(\mathbf{m}_{i}\cdot\mathbf{u}_{ij}\right)^{2}, & i\in\mathrm{surface}.
\end{array}\right.\label{eq:HamUA-NSA}
\end{equation}

The macroscopic state of the NM may be described using what is often
called the \emph{superspin} or\emph{ macrospin}, that is the net magnetic
moment
\begin{equation}
\mathbf{m}=\frac{1}{\mathcal{N}}\sum_{i=1}^{\mathcal{N}}\mathbf{m}_{i}.\label{eq:Macrospin}
\end{equation}

The dynamics of the magnetic moments $\mathbf{m}_{i}$ is governed
by the (damped) Landau-Lifshitz equation (LLE) or, more precisely,
the system of coupled Landau-Lifshitz equations written for the atomic
magnetic moments $\mathbf{m}_{i}$ ($i=1,2,\ldots,\mathcal{N}$),
\begin{equation}
\frac{d\mathbf{m}_{i}}{d\tau}=\mathbf{m}_{i}\times\mathbf{h}_{{\rm eff},i}-\alpha\mathbf{m}_{i}\times\left(\mathbf{m}_{i}\times\mathbf{h}_{{\rm eff},i}\right),\label{eq:LLE-MSP}
\end{equation}
with the (normalized) local effective field $\mathbf{h}_{{\rm eff},i}$,
acting on $\mathbf{s}_{i}$, being defined by $\mathbf{h}_{{\rm eff},i}=-\delta\mathcal{H}/\delta\mathbf{s}_{i}$.
$\tau$ is the reduced time given by $\tau=t/\tau_{{\rm s}},$ where
$\tau_{{\rm s}}=\mu_{a}/\left(\gamma J\right)$ is a characteristic
time of the system's dynamics; $\gamma\simeq1.76\times10^{11}$ (Ts)$^{-1}$
is the gyromagnetic ratio and $\alpha$ the damping parameter ($\sim0.01-0.1$).
For example, for cobalt $J=8{\rm \ meV}$ and $\tau_{s}=70\ {\rm fs}$.
In these units, $\mathbf{h}_{{\rm eff},i}=\mu_{a}\mathbf{H}_{{\rm eff},i}/J$,
where $\mathbf{H}_{i}^{{\rm eff}}=-\left(1/\mu_{a}\right)\left(\delta\mathcal{H}/\delta\mathbf{m}_{i}\right)$
is the deterministic field that comprises the exchange field, the
magnetic field $\mathbf{H}_{\mathrm{ext}}$ and the anisotropy field
$\mathbf{H}_{A}$.

\begin{figure}[h!]
\begin{centering}
\includegraphics[angle=-45,width=5.5cm]{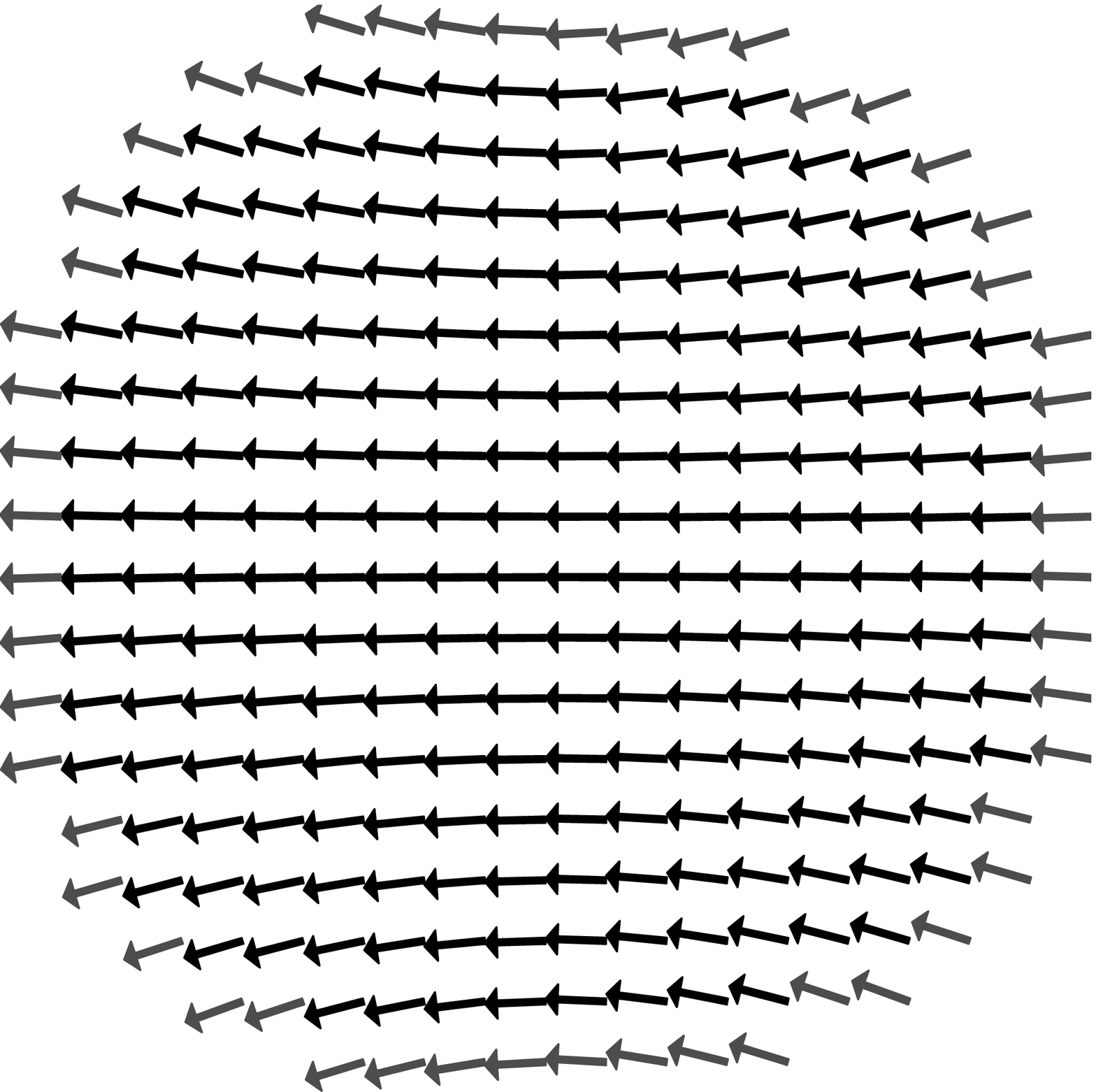}
\par\end{centering}
\caption{\label{fig:nsa-structure} Magnetic structure of a spherical nanoparticle
of linear size $N=20$, showing atoms in the plane $z=0$.}
\end{figure}

The spin configuration shown in Fig. \ref{fig:nsa-structure}, with
the net magnetic moment along the diagonal, is obtained by (numerically)
minimizing the energy (\ref{eq:Ham-MSP}) by solving the system of
coupled Landau-Lifshitz equations (\ref{eq:LLE-MSP}) \cite{kacdim02prb,garkac03prl,kacmah04jmmm,kacgar05springer,kacbon06prb,kachkachi07j3m}.
This is a typical spin structure that is induced by the NSA in a spherical
NM. Note that the atomic magnetic moments $\mathbf{m}_{i}$ progressively
deviate from the global orientation (here the diagonal) as the site
$i$ is located closer to the NM border and away from the diagonal.

\noindent \emph{Orders of magnitude of materials parameters:} Let
us now give a few orders of magnitude of the physical parameters that
appear in the Hamiltonian (\ref{eq:Ham-MSP}). First of all, we note
that Eq. (\ref{eq:Ham-MSP}) is the energy per atom, obtained by dividing
the total energy of the system by $\mathcal{\ensuremath{N}}$, the
number of atoms in the NM. Hence, the physical parameters involved,
namely $J,K$ and $\mu_{a}\left(\mu_{0}H\right)$, are measured in
Joule per atom. For instance, the anisotropy energy, which is often
written as $\tilde{K}V$ where $V$ is the volume of the NM and $\tilde{K}$
the density of anisotropy energy (in ${\rm J/m}^{3}$), becomes $\tilde{K}V=\mathcal{\ensuremath{N}}v_{0}\tilde{K}\equiv\mathcal{\ensuremath{N}}K$.
Similarly, the Zeeman contribution which usually reads $\mu_{0}HM$
is now rewritten as $\mu_{0}HM=\mathcal{\ensuremath{N}}\mu_{a}\left(\mu_{0}H\right)$
\footnote{The magnetic field $H$ in (\ref{eq:Ham-MSP}), and all subsequent
equations, should be understood as $\mu_{0}H$ which is measured in
Tesla, so that the Zeeman term $\mu_{a}\left(\mu_{0}H\right)$ is
measured in J/atom.}. In the NSA model, we distinguish between the core (full coordination)
and surface atoms (with smaller coordination). As such, the anisotropy
constant $K_{c}$ applies only to atoms in the core of the NM and
$K_{s}$ only to those on its surface. For instance, for cobalt, the
magnetic moment per atom $\mu_{a}=n_{0}\mu_{B}$, with $n_{0}$ being
the number of Bohr magnetons per atom ($n_{0}\simeq1.7$) and $\mu_{B}=9.274\times10^{-24}$J/T
is the Bohr magneton. Hence, $\mu_{a}\simeq1.58\times10^{-23}\,{\rm J/T}$.
Next, the magneto-crystalline anisotropy constant is roughly $K_{c}\simeq3\times10^{-24}$
J/atom, the surface anisotropy constant is around $K_{s}\simeq5.22\times10^{-23}\,$
Joule/atom and the (bulk) exchange coupling is $J\simeq8\,\mathrm{mev}$
or $1.2834\times10^{-21}$ J/atom. The lattice parameter is $a=0.3554$
nm. As such, $k_{c}\equiv K_{c}/J\simeq0.00234$ while $k_{s}\equiv K_{s}/J\simeq0.04$.
The latter value is within the range of values estimated by several
experimental studies. Indeed, one may find $K_{s}/J\simeq0.1$ for
cobalt \cite{skocoe99iop}, $K_{s}/J\simeq0.06$ for iron \cite{urquhartetal88jap},
and $K_{s}/J\simeq0.04$ for maghemite particles \cite{perrai05springer}.

\subsection{Continuum approach}

In the continuum approach, the magnetic configuration of a system
is described by the continuous magnetization vector field $\mathbf{M}\left(\mathbf{r}\right)$
constrained to a constant norm $M_{s}$. The relation between the
discrete and continuous descriptions is \cite{Turov65ap}
\begin{equation}
\mathbf{M}\left(\mathbf{r}\right)=\sum_{i}\boldsymbol{\mu}_{i}/v_{0}\label{eq:Mr}
\end{equation}
where $\boldsymbol{\mu}_{i}$ is the discrete magnetic moment of the
$i^{\mathrm{th}}$ ion belonging to a given sub-lattice. The summation
is carried over all sites in a physically small volume $v_{0}$, around
a point whose position is $\mathbf{r}=(x,y,z)$, and within which
the moments $\boldsymbol{\mu}_{i}$ are assumed to be uniform. The
normalized magnetization density vector field is then defined by
\begin{equation}
\mathbf{m}(\mathbf{r})\equiv\mathbf{M}\left(\mathbf{r}\right)/M_{s}.\label{eq:mr}
\end{equation}

In the continuum limit, the exchange interaction is written in terms
of the exchange stiffness $A$ (\emph{e.g.} about $3.6\,{\rm pJ/m}$
for cobalt) as an integral over the volume $V$ of the NM

\begin{eqnarray}
\mathcal{H}_{{\rm exc}} & =A\int_{V}d\mathbf{r}\,\sum_{\alpha=x,y,z}\left(\mathbf{\nabla}m_{\alpha}\cdot\mathbf{\nabla}m_{\alpha}\right).\label{eq:HexCL-m}
\end{eqnarray}

For a simple cubic lattice, we have the relation between $J$ and
$A$: $A=J/2a$, with $a$ being the lattice constant. This is the
classical analog of the relation that applies to a simple cubic lattice
of quantum spins, $A=J\left\langle S^{2}\right\rangle /2a$.

Using the identity $\nabla\cdot(m_{\alpha}\cdot\nabla m_{\alpha})=m_{\alpha}\Delta m_{\alpha}+(\nabla m_{\alpha})^{2}$
and the divergence theorem, the exchange energy can split into a core
and a surface contribution, namely
\begin{eqnarray}
\mathcal{H}_{{\rm exc}} & =-A\sum_{\alpha=x,y,z}\int_{V}m_{\alpha}\Delta m_{\alpha}\;\mathrm{d}^{3}r+A\sum_{\alpha=x,y,z}\oint_{\partial V}m_{\alpha}\nabla m_{\alpha}\;\mathbf{n}\,\mathrm{d}^{2}r.\label{eq:HexCL-m-2}
\end{eqnarray}

Next, the Zeeman term reads,
\begin{eqnarray}
\mathcal{H}_{{\rm Z}} & =-M_{s}\mathbf{H}_{\mathrm{ext}}\cdot\int_{V}\mathrm{d}^{3}r\,\mathbf{m}\left(\mathbf{r}\right)\label{eq:Zeeman-cont}
\end{eqnarray}
and the anisotropy energy for core spins becomes
\begin{eqnarray*}
\mathcal{H}_{{\rm core}} & =-\frac{K_{c}}{v_{0}}\int_{V_{{\rm in}}}d^{3}r\,\left(\mathbf{m}\cdot\mathbf{e}_{z}\right)^{2}.
\end{eqnarray*}
Note that $V_{{\rm in}}/v_{0}$, with $v_{0}=a^{3}$, is equal to
the number of core atoms that we denote by $N_{{\rm c}}$, so that
$V_{{\rm in}}K_{c}/v_{0}=N_{{\rm c}}K_{c}$. Regarding surface anisotropy,
it was shown in Ref. \cite{garkac03prl} that the corresponding energy
in the NSA model can be replaced by the approximate expression for
a sphere {[}see Ref. \cite{garanin18prb} for a cube{]}

\begin{equation}
\mathcal{H}_{{\rm surface}}=-\frac{K_{s}}{2}\sum_{\alpha}\left|n_{\alpha}\right|m_{\alpha}^{2},\label{eq:ESNeel-ave}
\end{equation}
where $\mathbf{n}$ is the unit vector of the normal to the surface
(the boundary $\partial V$ of the NM). Then, in the continuum limit,
$\mathcal{H}_{{\rm surface}}$ becomes
\[
\mathcal{H}_{{\rm surface}}=-\frac{K_{s}}{2a^{2}}\sum_{\alpha=x,y,z}\oint_{\partial V}d^{2}r\left|n_{\alpha}\right|m_{\alpha}^{2}.
\]

Therefore, collecting all contributions, the nanomagnet's Hamiltonian in Eq. (\ref{eq:Ham-MSP}) becomes in the continuum approach
\begin{eqnarray}
\mathcal{H} & =A\sum_{\alpha\in\{x,y,z\}}\left[\oint_{\partial V}m_{\alpha}\mathbf{\nabla}m_{\alpha}\cdot\mathbf{n}\,\mathrm{d}^{2}r-\int_{V}m_{\alpha}\Delta m_{\alpha}\;\mathrm{d}^{3}r\right]\label{eq:Ham-MSP-continuum}\\
 & -M_{s}\mathbf{H}_{\mathrm{ext}}\cdot\int_{V}d^{3}r\,\mathbf{m}\left(\mathbf{r}\right)\nonumber \\
 & -\frac{K_{c}}{a^{3}}\int_{V}\left(\mathbf{m}\cdot\mathbf{e}_{A}\right)^{2}\;\mathrm{d}^{3}r-\frac{K_{s}}{2a^{2}}\sum_{\alpha\in\{x,y,z\}}\oint_{\partial V}\left|n_{\alpha}\right|m_{\alpha}^{2}\;\mathrm{d}^{2}r.\nonumber
\end{eqnarray}

In the case of small spin-misalignment, where the magnetization density
$\mathbf{m}\left(\mathbf{r}\right)$ slightly deviates from the homogeneous
magnetization state $\mathbf{m}_{0}$ {[}see Fig. \ref{fig:nsa-structure}{]},
a perturbation approach is applicable. Accordingly, $\mathbf{m}_{0}$
is considered as the principal unit vector \footnote{In all subsequent formulae, $\mathbf{m}_{0}$ is considered as a known uniform vector field.} associated with $\mathbf{m}(\mathbf{r})$ while the spin-misalignment
is encoded in the vector field $\boldsymbol{\psi}\left(\mathbf{r}\right)$
with $\boldsymbol{\psi}\left(\mathbf{r}\right)\perp\mathbf{m}_{0}$.
Therefore, we write \cite{poylakov75plb, garkac09prb}

\begin{equation}
\mathbf{m}\left(\mathbf{r}\right)=\mathbf{m}_{0}\sqrt{1-\|\boldsymbol{\psi}\left(\mathbf{r}\right)\|^{2}}+\boldsymbol{\psi}\left(\mathbf{r}\right),\label{eq:ExactLinarization}
\end{equation}
with $\mathbf{m}_{0}\cdot\mathbf{\boldsymbol{\psi}}=0$ and thereby $\left|\mathbf{m}\left(\mathbf{r}\right)\right|=1$,
together with the condition (discussed later in the text)
\begin{equation}
\int_{V}d^{3}r\,\mathbf{\boldsymbol{\psi}}\left(\mathbf{r}\right)=0.\label{eq:Intpsizero}
\end{equation}

Assuming that $\psi_{\alpha}\ll1,\alpha=x,y,z$, an approximate closed-form
solution for the normalized magnetization density can be obtained
by performing the second-order expansion of equation Eq. (\ref{eq:ExactLinarization})
with respect to $\boldsymbol{\psi}$ :
\begin{equation}
\mathbf{m}\left(\mathbf{r}\right)\cong\mathbf{m}_{0}+\boldsymbol{\psi}\left(\mathbf{r}\right)-\frac{1}{2}\|\boldsymbol{\psi}\left(\mathbf{r}\right)\|^{2}\mathbf{m}_{0}.\label{eq:SecondOrderApproximation}
\end{equation}

Next, one can rewrite the Hamiltonian (\ref{eq:Ham-MSP-continuum})
using a perturbation approach and minimizing with respect to the Cartesian
components $\psi_{x},\psi_{y},\psi_{z}$, leading to a homogeneous
(vector) Helmholtz equation for these components, together with inhomogeneous
Neumann boundary conditions. However, owing to the transverse character
of $\mathbf{\boldsymbol{\psi}}$ ($\mathbf{m}_{0}\cdot\mathbf{\boldsymbol{\psi}}=0$),
it is more convenient to work in the local frame attached to $\mathbf{m}_{0}$,
\emph{i.e. }$\left(\mathbf{m}_{0},\mathbf{u}_{1},\mathbf{u}_{2}\right)$,
where $\mathbf{u}_{1}$ and $\mathbf{u}_{2}$ are the following two
unit vectors \cite{garkac09prb}
\begin{eqnarray}
\mathbf{u}_{1} & = & \frac{\mathbf{m}_{0}\times\mathbf{e}_{A}}{\left\Vert \mathbf{m}_{0}\times\mathbf{e}_{A}\right\Vert },\quad\mathbf{u}_{2}=\frac{\mathbf{m}_{0}\left(\mathbf{m}_{0}\cdot\mathbf{e}_{A}\right)-\mathbf{e}_{A}}{\left\Vert \mathbf{m}_{0}\left(\mathbf{m}_{0}\cdot\mathbf{e}_{A}\right)-\mathbf{e}_{A}\right\Vert }.\label{eq:u1u2-unitvectors}
\end{eqnarray}
In the new frame, we have
\begin{equation}
\mathbf{\boldsymbol{\psi}}=\psi_{1}\mathbf{u}_{1}+\psi_{2}\mathbf{u}_{2}\label{eq:Parametrization-12}
\end{equation}
and $\|\boldsymbol{\psi}\left(\mathbf{r}\right)\|^{2}=\psi_{1}^{2}+\psi_{2}^{2}$.
Consequently, with the help of a linear transformation, the problem
is readily reduced to the following system of decoupled homogeneous
scalar (dimensionless) Helmholtz equations
\begin{equation}
\left[\Delta_{\xi}-\kappa_{\beta}^{2}\right]\psi_{\beta}\left(\boldsymbol{\xi}\right)=0,\quad\beta=1,2\label{eq:DecoupledHelmholtzEquation}
\end{equation}
along with the inhomogeneous Neumann boundary conditions
\begin{equation}
\left.\frac{\mathrm{d}\psi_{\beta}}{\mathrm{d}\xi}\right|_{\xi=1}=\widetilde{k}_{s}\sum_{\alpha=x,y,z}\left|n_{\alpha}\right|\left(\mathbf{m}_{0}\cdot\mathbf{e}_{\alpha}\right)\left(\mathbf{u}_{\beta}\cdot\mathbf{e}_{\alpha}\right),\quad\beta=1,2.\label{eq:NeumannBC}
\end{equation}
Here we have introduced the dimensionless coordinates $\boldsymbol{\xi}=\mathbf{r}/R$,
where $R$ is the NM radius, together with the following Helmholtz
coefficients $\kappa_{\beta},\beta=1,2$, given by
\begin{equation}
\left\{ \begin{array}{lll}
\kappa_{1}^{2} & = & \left(\widetilde{\mathbf{h}}_{\mathrm{ext}}\cdot\mathbf{m}_{0}\right)+2\widetilde{k}_{c}\cdot\left(\mathbf{m}_{0}\cdot\mathbf{e}_{A}\right)^{2},\\
\\
\kappa_{2}^{2} & = & \left(\widetilde{\mathbf{h}}_{\mathrm{ext}}\cdot\mathbf{m}_{0}\right)+2\widetilde{k}_{c}\cdot\left[2\left(\mathbf{m}_{0}\cdot\mathbf{e}_{A}\right)^{2}-1\right].
\end{array}\right.\label{eq:HelmholtzCoefficients}
\end{equation}
where

\begin{equation}
\tilde{k}_{c} =\frac{1}{2}\left(\frac{D}{a}\right)^{2}k_{c}, \qquad \tilde{k}_{s} =\left(\frac{D}{a}\right)k_{s}, \qquad
\tilde{h}_{\mathrm{ext}} =\frac{1}{2}\left(\frac{D}{a}\right)^{2}h_{\mathrm{ext}},\label{eq:DecoupledHelmholtzEquationCoefficient3}
\end{equation}
with $D=2R$ being the diameter of the NM, $k_{c}\equiv K_{c}/J$
and $k_{s}\equiv K_{s}/J$ the (dimensionless) reduced anisotropy
constants introduced earlier, and $h_{\mathrm{ext}}\equiv\mu_{a}H_{\mathrm{ext}}/J$
the reduced magnetic field.

For later use and simplicity of notation, we introduce the surface anisotropy field
\begin{equation}
\Sigma_{\beta}\left(\mathbf{m}_{0},\mathbf{n}\right)\equiv\left.\frac{\mathrm{d}\psi_{\beta}}{\mathrm{d}\xi}\right|_{\xi=1}.\label{eq:SAF}
\end{equation}

In the case of a core anisotropy easy axis in the $z$ direction,
$\mathbf{e}_{A}=\mathbf{e}_{z}$,
\begin{eqnarray}
\mathbf{u}_{1} & = & \frac{\mathbf{m}_{0}\times\mathbf{e}_{z}}{\sqrt{1-m_{0,z}^{2}}},\quad\mathbf{u}_{2}=\frac{\mathbf{m}_{0}\left(\mathbf{m}_{0}\cdot\mathbf{e}_{z}\right)-\mathbf{e}_{z}}{\sqrt{1-m_{0,z}^{2}}},\label{eq:u1u2}
\end{eqnarray}
and
\begin{eqnarray}
\Sigma_{1} & =\widetilde{k}_{s}\frac{m_{0,x}m_{0,y}}{\sqrt{1-m_{0,z}^{2}}}\left(\left|n_{x}\right|-\left|n_{y}\right|\right),\label{eq:SAF-Comps}\\
\Sigma_{2} & =\widetilde{k}_{s}\frac{m_{0,z}}{\sqrt{1-m_{0,z}^{2}}}\left[\left(\left|n_{x}\right|-\left|n_{z}\right|\right)m_{0,x}^{2}+\left(\left|n_{y}\right|-\left|n_{z}\right|\right)m_{0,y}^{2}\right].\nonumber
\end{eqnarray}

\section{\label{sec:Magnetization-profile:GF}Magnetization profile: Green's
function approach}

To solve the homogeneous Helmholtz equation (\ref{eq:DecoupledHelmholtzEquation})
for $\psi_{\beta}$, with the inhomogeneous Neumann boundary conditions
(\ref{eq:NeumannBC}), a specified gradient on the surface, we use
the Green's function (GF) approach \cite{duffy2015green,morse1954methods,MorseFeschbach_mgh53}.
The GF $\mathcal{G}_{\beta}\left(\boldsymbol{\xi},\boldsymbol{\xi}'\right)$
for this problem satisfies the equation

\begin{equation}
\left[\Delta_{\xi}-\kappa_{\beta}^{2}\right]\mathcal{G}_{\beta}\left(\boldsymbol{\xi},\boldsymbol{\xi}'\right)=-4\pi\delta\left(\boldsymbol{\xi}-\boldsymbol{\xi}'\right),\label{eq:GF_Helmholtz_general}
\end{equation}
and may be chosen to satisfy the homogeneous boundary condition of
the same type as $\psi_{\beta}$, \emph{i.e.} Neumann boundary conditions,
\begin{equation}
\left.\frac{\mathrm{d}\mathcal{G}_{\beta}}{\mathrm{d}\xi}\right|_{\boldsymbol{\xi}=1}=0.\label{eq:g-BC}
\end{equation}

In this case, we have the solution \cite{MorseFeschbach_mgh53,morse1954methods,garkac03prl,duffy2015green}
\begin{equation}
\psi_{\beta}\left(\boldsymbol{\xi}\right)=\frac{1}{4\pi}\oint_{\partial V}d^{2}n^{\prime}\,\Sigma_{\beta}\left(\mathbf{m}_{0},\mathbf{n}^{\prime}\right)\mathcal{G}_{\beta}(\boldsymbol{\xi},\mathbf{n}')\label{eq:GF_Helmholtz_general3}
\end{equation}
for $\boldsymbol{\xi}$ inside and on its boundary $\partial V$,
where $\Sigma_{\beta}\left(\mathbf{m}_{0},\mathbf{n}\right)=\left.\frac{\mathrm{d}\psi_{\beta}}{\mathrm{d}\xi}\right|_{\boldsymbol{\xi}=\mathbf{n}}=\left(\mathbf{\nabla}_{\xi}\psi_{\beta}\right)\cdot\mathbf{n}$
is the outward normal gradient of $\psi_{\beta}$ at the surface of
the NM, with $\mathbf{n}=\left(\sin\theta\cos\varphi,\sin\theta\sin\varphi,\cos\theta\right)$
and $d^{2}n=d\Omega=\sin\theta\,d\theta d\varphi$.

The result in Eq. (\ref{eq:GF_Helmholtz_general3}) simply reflects
the fact that the source of spin mis-alignment $\psi_{\beta}$ within
the NM spin configuration is induced by surface anisotropy via the
field $\Sigma_{\beta}=\left.\frac{\mathrm{d}\psi_{\beta}}{\mathrm{d}\xi}\right|_{\boldsymbol{\xi}=\mathbf{n}}$
given in Eq. (\ref{eq:NeumannBC}). As discussed in Ref. \cite{garkac03prl},
the spin mis-alignment (or disorder) initiated at the surface of the
NM propagates into the body of the latter down to its center. In this
case, the contribution of surface anisotropy to the overall anisotropy
of the NM scales with its volume ($N^{3}$). In the presence of uniaxial
anisotropy in the core, the surface spin disorder is screened out
at a certain distance from the center and the contribution of the
surface to the overall anisotropy then scales as the surface ($N^{2}$)
{[}see Section \ref{subsec:In-the-presence-of-Kc} for further discussion{]}.

\subsection{\label{subsecKcZero}No core anisotropy}

In the absence of core anisotropy ($K_{{\rm c}}=0$) and magnetic
field, $\kappa_{1}^{2}=\kappa_{2}^{2}=0$ [see Eq. (\ref{eq:HelmholtzCoefficients})],
the vector $\mathbf{m}_{0}$ is along the cube diagonal, \textit{i.e.},
$m_{0,\alpha}=1/\sqrt{3}$. Then, Eq. (\ref{eq:DecoupledHelmholtzEquation})
reduces to the Laplace equation $\Delta\psi_{\beta}=0$, subjected
again to the inhomogeneous Neumann boundary conditions (\ref{eq:NeumannBC}).
The corresponding GF $\mathcal{G}^{\left(0\right)}\left(\boldsymbol{\xi},\boldsymbol{\xi}'\right)$
satisfies the Poisson equation
\begin{eqnarray}
\Delta\mathcal{G}^{\left(0\right)}\left(\boldsymbol{\xi},\boldsymbol{\xi}'\right) & =-4\pi\delta\left(\boldsymbol{\xi}-\boldsymbol{\xi}'\right).\label{eq:0thOrderGF}
\end{eqnarray}

However, integrating over the volume of the NM, we can see that the
homogeneous boundary condition (\ref{eq:g-BC}) can no longer be used.
Instead, setting $\left.\frac{\mathrm{d}\mathcal{G}_{\beta}}{\mathrm{d}\xi}\right|_{\boldsymbol{\xi}=1}=C$,
\emph{i.e.} an inhomogeneous Neumann boundary condition, one finds
that $C=-1$, and the GF function of the problem is then given by
\cite{garkac03prl,sadybekov2016construction} (up to a constant)
%
\begin{eqnarray}
\mathcal{G}^{(0)}\left(\boldsymbol{\xi},\boldsymbol{\xi}'\right) & =\frac{1}{\left|\boldsymbol{\xi}-\boldsymbol{\xi}'\right|}
+\frac{1}{\sqrt{1+\mathbf{\xi}^{2}\mathbf{\xi}^{\prime2}-2\left(\boldsymbol{\xi}\cdot\boldsymbol{\xi}'\right)}} \label{eq:GF0-rrp}\\
& \quad -\ln\left|1-\left(\boldsymbol{\xi}\cdot\boldsymbol{\xi}'\right)+\sqrt{1+\mathbf{\xi}^{2}\mathbf{\xi}^{\prime2}-2\left(\boldsymbol{\xi}\cdot\boldsymbol{\xi}'\right)}\right|. \nonumber\\
\left.\frac{\mathrm{d}\mathcal{G}^{(0)}}{\mathrm{d}\xi}\right|_{\xi=1} & =-1.\label{eq:g0-BC}
\end{eqnarray}

When one of the arguments is on the surface, \emph{i.e.} $\mathbf{\mathbf{\xi}}^{\prime}=\mathbf{n}^{\prime}$ ($\xi^{\prime}=1$), (\ref{eq:GF0-rrp}) simplifies into

\begin{eqnarray}
\mathcal{G}^{(0)}\left(\boldsymbol{\xi},\mathbf{n}^{\prime}\right) & =\frac{2}{\sqrt{1+\boldsymbol{\xi}^{2}-2\left(\boldsymbol{\xi}\cdot\mathbf{n}^{\prime}\right)}}\label{eq:GFn-vs}\\
&\quad-\ln\left|1-\left(\boldsymbol{\xi}\cdot\mathbf{n}^{\prime}\right)+\sqrt{1+\boldsymbol{\xi}^{2}-2\left(\boldsymbol{\xi}\cdot\mathbf{n}^{\prime}\right)}\right|.\nonumber
\end{eqnarray}

Note that, in general, because of the inhomogeneous boundary condition
(\ref{eq:g0-BC}), the GF $\mathcal{G}^{(0)}\left(\boldsymbol{\xi},\boldsymbol{\xi}'\right)$
looses its symmetry of interchange $\boldsymbol{\xi}\longleftrightarrow\boldsymbol{\xi}'$.
However, by imposing the condition $\oint_{\partial V}\mathcal{G}^{(0)}(\boldsymbol{\xi},\mathbf{n}')\,d^{2}n'=0$,
this symmetry is restored. The GF $\mathcal{G}^{(0)}\left(\boldsymbol{\xi},\boldsymbol{\xi}'\right)$
in (\ref{eq:GFn-vs}) yields $\left(1/4\pi\right)\oint_{\partial V}\mathcal{G}^{(0)}(\boldsymbol{\xi},\mathbf{n}')\,d^{2}n'=2-\ln2$,
and hence by making the replacement $\mathcal{G}^{(0)}\rightarrow\mathcal{G}^{(0)}-\left(2-\ln2\right)$,
which does not modify the boundary condition (\ref{eq:g0-BC}), we
restore the symmetry $\boldsymbol{\xi}\longleftrightarrow\boldsymbol{\xi}'$.

For $\xi\ll1$ (or $r\ll R$), we obtain the fourth-order expansion
\begin{eqnarray}
\mathcal{G}^{(0)}\left(\boldsymbol{\xi},\mathbf{n}\right) & \simeq2-\ln2+3\left(\mathbf{n}\cdot\boldsymbol{\xi}\right)+\frac{5}{4}\left[3\left(\mathbf{n}\cdot\boldsymbol{\xi}\right)^{2}-\xi^{2}\right]\nonumber \\
 & +\frac{7}{6}\left(\mathbf{n}\cdot\boldsymbol{\xi}\right)\left[5\left(\mathbf{n}\cdot\boldsymbol{\xi}\right)^{2}-3\xi^{2}\right]\label{eq:GF0-Expansion}\\
 & +\frac{9}{32}\left[35\left(\mathbf{n}\cdot\boldsymbol{\xi}\right)^{4}-30\left(\mathbf{n}\cdot\boldsymbol{\xi}\right)^{2}\xi^{2}+3\xi^{4}\right]+\ldots\nonumber
\end{eqnarray}
In this Section ($K_{{\rm c}}=0$), the spin deviation $\psi_{\beta}$
is then given by \cite{MorseFeschbach_mgh53}
\begin{equation}
\psi_{\beta}^{\left(0\right)}\left(\boldsymbol{\xi}\right)=\frac{1}{4\pi}\oint_{\partial V}d^{2}n^{\prime}\,\Sigma_{\beta}\left(\mathbf{m}_{0},\mathbf{n}^{\prime}\right)\mathcal{G}^{\left(0\right)}\left(\boldsymbol{\xi},\mathbf{n}^{\prime}\right).\label{eq:psi-Coeffs0}
\end{equation}

Note that had we kept the constant $2-\ln2$ in $\mathcal{G}^{(0)}\left(\boldsymbol{\xi},\mathbf{n}\right)$,
we would have obtained the same result since the contribution of this
constant term vanishes under the surface integral when we substitute
$\Sigma_{\beta}$ from Eqs. (\ref{eq:SAF}, \ref{eq:NeumannBC}).
For the same reason, odd-order terms in the expansion (\ref{eq:GF0-Expansion})
do not contribute to (\ref{eq:psi-Coeffs0}).

Therefore, using (\ref{eq:NeumannBC}) and the expansion (\ref{eq:GF0-Expansion})
up to $4^{\mathrm{nd}}$ order in $\xi$, we obtain the following
explicit expressions ($\xi=r/R$) for the components of the spin deviation
vector $\boldsymbol{\psi}^{\left(0\right)}$:
\begin{eqnarray}
&\psi_{1}^{\left(0\right)} \simeq\lambda_{s}\frac{m_{0,x}m_{0,y}}{\sqrt{1-m_{0,z}^{2}}}\left(\xi_{x}^{2}-\xi_{y}^{2}\right)\left[1+\frac{1}{16}\left(7\xi_{z}^{2}-\xi^{2}\right)\right],\label{eq:psi0-12-sol}\\
&\psi_{2}^{\left(0\right)} \simeq\lambda_{s}\frac{m_{0,z}}{\sqrt{1-m_{0,z}^{2}}}\left\{
\begin{array}{l}
m_{0,x}^{2}\left(\xi_{x}^{2}-\xi_{z}^{2}\right)+m_{0,y}^{2}\left(\xi_{y}^{2}-\xi_{z}^{2}\right)+\\
\frac{1}{16}\left[m_{0,x}^{2}\left(\xi_{x}^{2}-\xi_{z}^{2}\right)\left(7\xi_{y}^{2}-\xi^{2}\right)+m_{0,y}^{2}\left(\xi_{y}^{2}-\xi_{z}^{2}\right)\left(7\xi_{x}^{2}-\xi^{2}\right)\right]
\end{array}\right\} ,\nonumber
\end{eqnarray}
where $\lambda_{s}\equiv15\widetilde{k}_{s}/32$. These expressions
have been obtained using the following integrals
\begin{eqnarray}
&&\frac{3}{4\pi}\oint_{\partial V}\left(\mathbf{n}\cdot\boldsymbol{\xi}\right)\left|n_{\alpha}\right|\;\mathrm{d}^{2}n =0,\quad\alpha=x,y,z,\label{eq:SomeIntegs}\\
&& \frac{1}{4\pi}\oint_{\partial V}\left[3\left(\mathbf{n}\cdot\boldsymbol{\xi}\right)^{2}-\xi^{2}\right]\left|n_{\alpha}\right|\;\mathrm{d}^{2}n =\frac{1}{8}\left(3\xi_{\alpha}^{2}-\xi^{2}\right),\nonumber \\
&& \frac{1}{4\pi}\oint_{\partial V}\left[35\left(\mathbf{n}\cdot\boldsymbol{\xi}\right)^{4}-30\left(\mathbf{n}\cdot\boldsymbol{\xi}\right)^{2}\xi^{2}+3\xi^{4}\right]\left|n_{\alpha}\right|\;\mathrm{d}^{2}n =-\frac{1}{48}\left(35\xi_{\alpha}^{4}-30\xi^{2}\xi_{\alpha}^{2}+3\xi^{4}\right).\nonumber
\end{eqnarray}

We see in Eq. (\ref{eq:psi0-12-sol}) that the spin deviation $\mathbf{\boldsymbol{\psi}}^{\left(0\right)}$,
caused by surface anisotropy, is linear in the corresponding constant
$k_{s}$ (through $\lambda_{s}$) and depends on the equilibrium magnetic
moment $\mathbf{m}_{0}$. It is also clear that this deviation depends
on the position within the NM and on the direction along which $\xi$
is varied from the center out to the boundary of the NM. These results
corroborate the discussion of the spin configuration shown in Fig.
\ref{fig:nsa-structure}.

When $\mathbf{r}$ is on the surface, \emph{i.e.} $\mathbf{\xi}=\mathbf{n}$,
we obtain the largest deviation with respect to the homogeneous state
$\mathbf{m}_{0}$ (using the $\xi^{2}$ expansion):

\begin{eqnarray}
\psi_{1}^{\left(0\right)}\left(\mathbf{n},\mathbf{m}_{0}\right) & \simeq\lambda_{s}\left(n_{x}^{2}-n_{y}^{2}\right)\frac{m_{0,x}m_{0,y}}{\sqrt{1-m_{0,z}^{2}}},\label{eq:psi0-LargeDev}\\
\psi_{2}^{\left(0\right)}\left(\mathbf{n},\mathbf{m}_{0}\right) & \simeq\lambda_{s}\frac{m_{0,z}}{\sqrt{1-m_{0,z}^{2}}}\left[\left(n_{x}^{2}-n_{z}^{2}\right)m_{0,x}^{2}+\left(n_{y}^{2}-n_{z}^{2}\right)m_{0,y}^{2}\right].\nonumber
\end{eqnarray}

\begin{figure*}[h!]
\begin{centering}
\includegraphics[width=7.5cm]{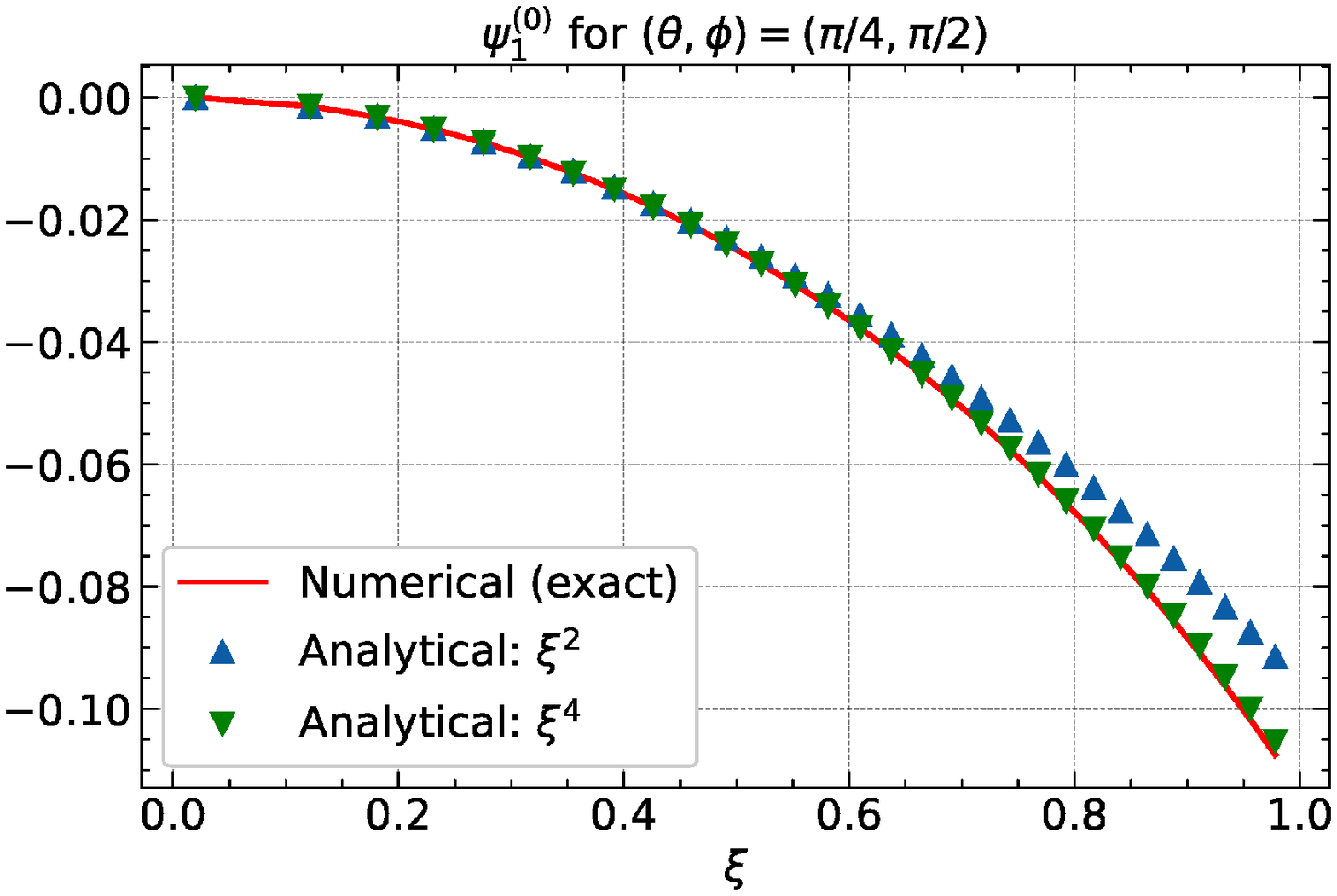} \includegraphics[width=7.5cm]{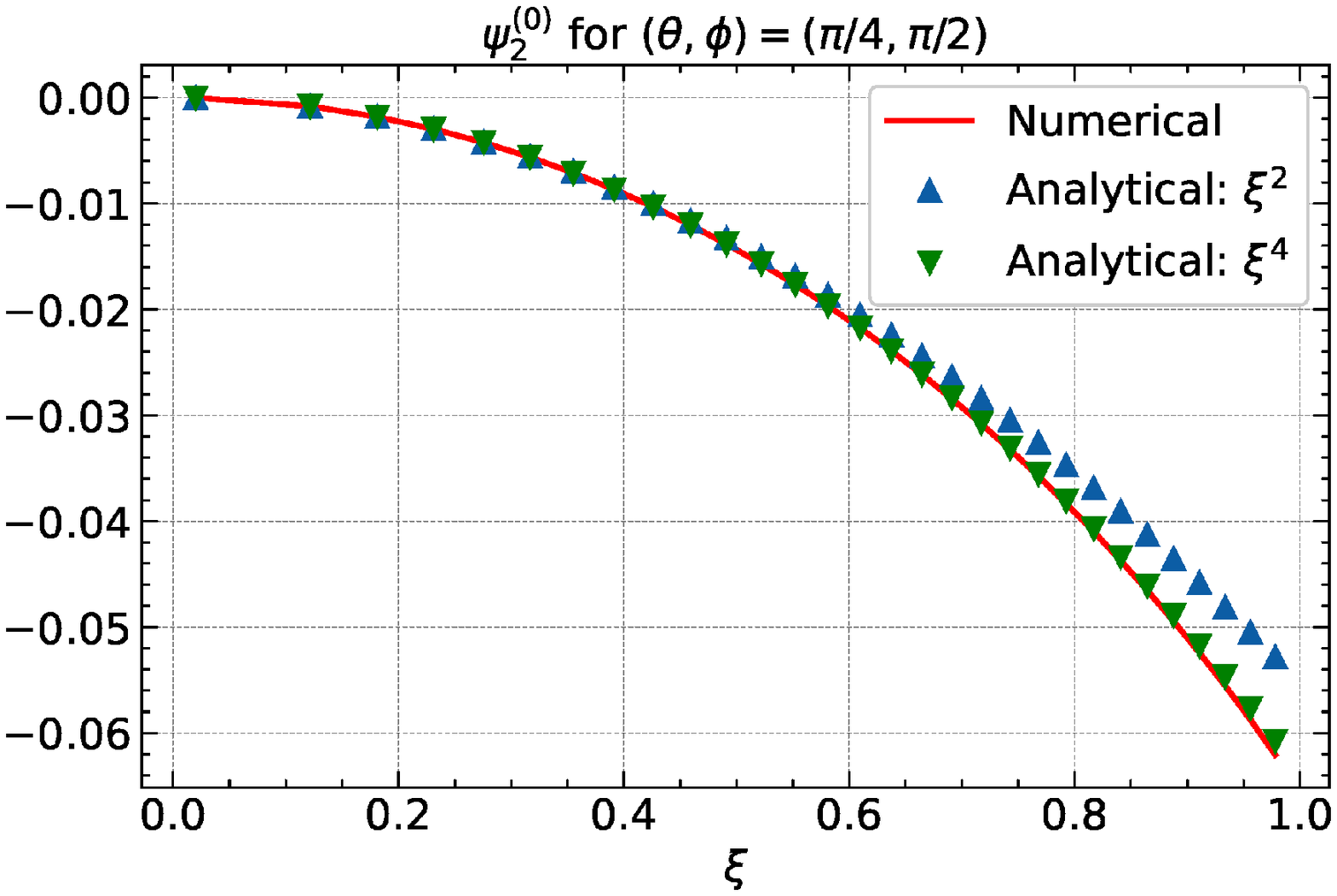}
\par\end{centering}
\caption{\label{fig:psi01psi02-exact-expanded} Components $\psi_{\alpha}^{(0)},\alpha=1,2$,
of the spin deviation vector $\boldsymbol{\psi}$ as a function of (scaled)
distance $\xi$ from the center of the nanomagnet. The red curve is
a plot of the integral (\ref{eq:psi-Coeffs0}), numerically computed
using the exact Green's function (\ref{eq:GFn-vs}). The blue and
green curves (in symbols) are, respectively, obtained from Eq. (\ref{eq:GF0-Expansion})
using the second and fourth order expansions. For both components,
the direction of $\mathbf{\xi}$ is set to $\theta=\pi/4,\varphi=\pi/2$.
$k_{c}=0,k_{s}=0.1$. {[}No magnetic field{]}.}
\end{figure*}

As we will see below and in the next Section, it is handier in practice
to use the expansions in Eq. (\ref{eq:psi0-12-sol}) than the exact
integral (\ref{eq:psi-Coeffs0}). For this purpose, we compare the
two in Fig. \ref{fig:psi01psi02-exact-expanded}.

The plots in Fig. \ref{fig:psi01psi02-exact-expanded} show that the
$4^{\mathrm{th}}-$order expansion of the Green function (\ref{eq:GFn-vs})
renders a fairly good approximation to the components of $\boldsymbol{\psi}$
for all $\xi$ between $0$ and $1$, \textit{i.e.} from the center
of the nanomagnet up to its border.

\begin{figure*}[h!]
\begin{centering}
\includegraphics[width=7.5cm]{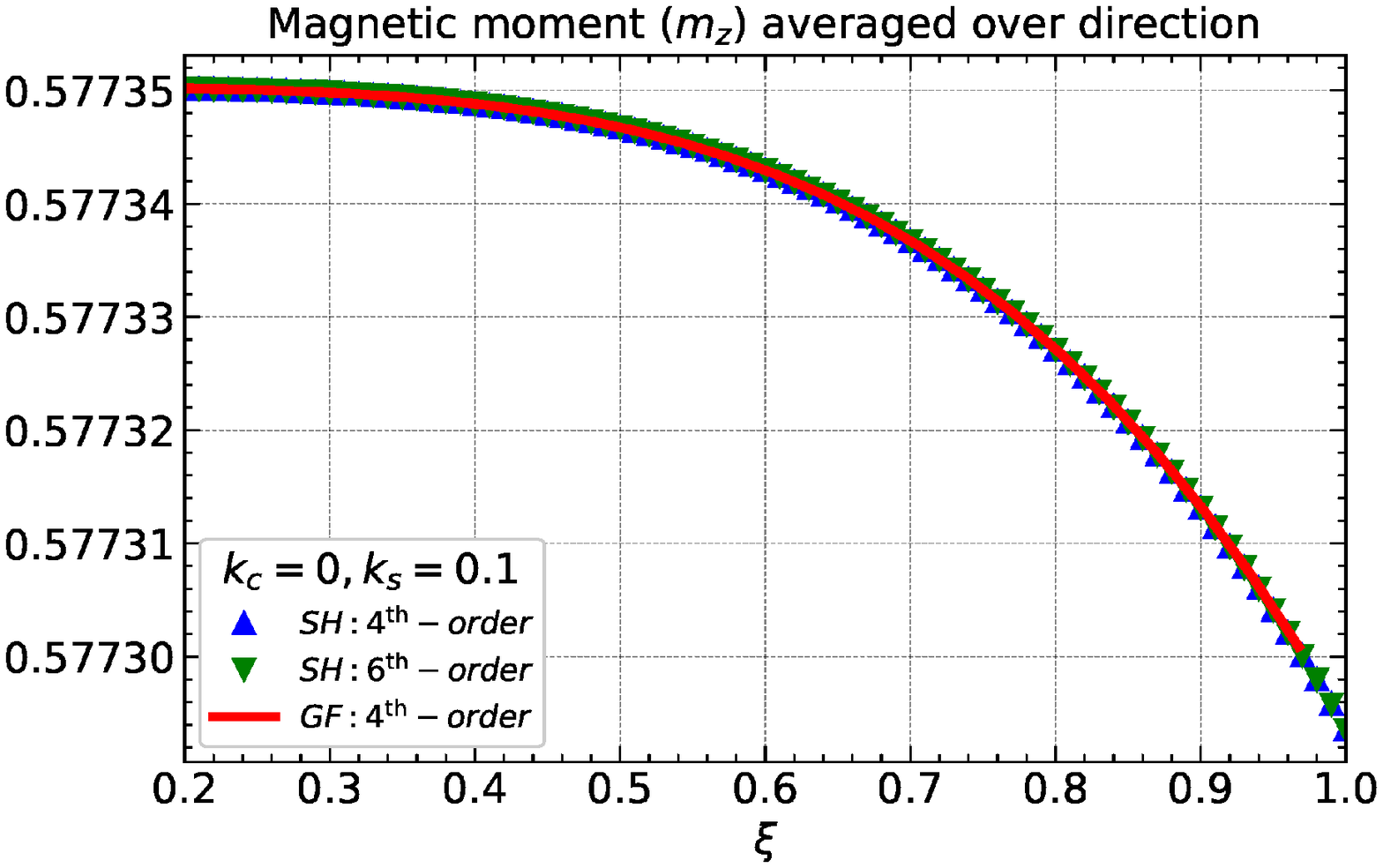} \includegraphics[width=7.5cm]{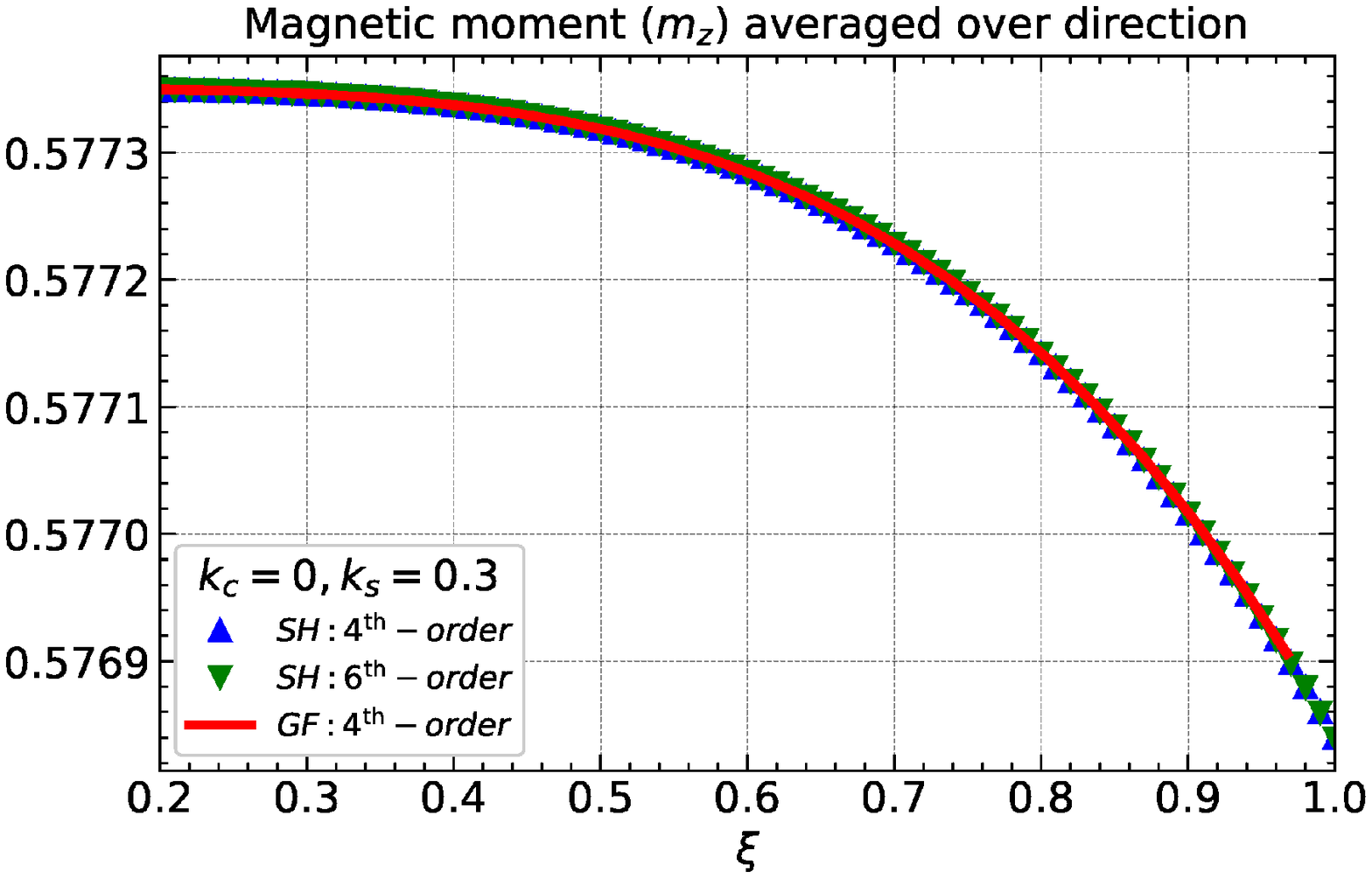}
\par\end{centering}
\caption{\label{fig:mx-GF-vs-SH} Component of the net magnetic moment, averaged
over the solid angle, as a function of $\xi$, for $k_{c}=0,k_{s}=0.1$
(left) and $k_{c}=0.0,k_{s}=0.3$ (right), as given by the GF approach
($4^{\mathrm{th}}-$order) and SH approach ($4^{\mathrm{th}}$ and
$6^{\mathrm{th}}$ orders). {[}No magnetic field{]}.}
\end{figure*}

We have also compared these results, rendered by the Green's function
technique, to the solution of the same boundary problem using the
technique of spherical harmonics (SH), presented in Refs. \cite{AdamsEtal_jac22_ana,AdamsEtal_jac22_num}.
The outcome of this comparison is shown in Fig. \ref{fig:mx-GF-vs-SH}
for two values of $k_{s}$. Note that here we have averaged the net
magnetic moment over the direction solid angle {[}see discussion below{]}.
We see that the $4^{\mathrm{th}}-$order approximation given in Eq.
(\ref{eq:GF0-Expansion}) and adopted here for the Green's function
$\mathcal{G}^{(0)}\left(\mathbf{n},\boldsymbol{\xi}^{\prime}\right)$,
agrees very well with the expansion in terms of spherical harmonics
up to the same order, to the $6^{\mathrm{th}}$ and even to the $10^{\mathrm{th}}$
order (not shown).

Next, using (\ref{eq:SecondOrderApproximation}), we can now write
explicit expressions for the components of the NM magnetic moment,
$m_{\alpha},\alpha=x,y,z$. In the frame $\left(\mathbf{m}_{0},\mathbf{u}_{1},\mathbf{u}_{2}\right)$
with $\mathbf{u}_{1}$ and $\mathbf{u}_{2}$ given in Eq. (\ref{eq:u1u2}),
we have

\[
\boldsymbol{\psi}\left(\mathbf{r}\right)=\psi_{1}\mathbf{u}_{1}+\psi_{2}\mathbf{u}_{2}
\]
or using $m_{0,\alpha}=1/\sqrt{3}$ and
\begin{eqnarray*}
\mathbf{u}_{1}\cdot\mathbf{e}_{x} & =\frac{1}{\sqrt{2}},\mathbf{u}_{1}\cdot\mathbf{e}_{y}=\frac{-1}{\sqrt{2}},\mathbf{u}_{1}\cdot\mathbf{e}_{z}=0,\\
\mathbf{u}_{2}\cdot\mathbf{e}_{x} & =\frac{1}{\sqrt{6}},\mathbf{u}_{2}\cdot\mathbf{e}_{y}=\frac{1}{\sqrt{6}},\mathbf{u}_{2}\cdot\mathbf{e}_{z}=-\sqrt{\frac{2}{3}},
\end{eqnarray*}
we write $\boldsymbol{\psi}\left(\mathbf{r}\right)$ in the frame
$\left(\mathbf{e}_{x},\mathbf{e}_{y},\mathbf{e}_{z}\right)$ as :

\begin{eqnarray*}
\boldsymbol{\psi}\left(\mathbf{r}\right) & =\psi_{x}\mathbf{e}_{x}+\psi_{y}\mathbf{e}_{y}+\psi_{z}\mathbf{e}_{z}=\frac{1}{\sqrt{2}}\left(\begin{array}{l}
\psi_{1}+\frac{1}{\sqrt{3}}\psi_{2}\\
-\psi_{1}+\frac{1}{\sqrt{3}}\psi_{2}\\
-\frac{2}{\sqrt{3}}\psi_{2}
\end{array}\right).
\end{eqnarray*}

Therefore, to $2^{\mathrm{nd}}$ order in $r$ (or $\xi$), we obtain
the spatial profile of the net magnetic moment (for $k_{c}=0$)
\begin{eqnarray*}
\mathbf{m}\left(\mathbf{r}\right) & \cong\mathbf{m}_{0}+\boldsymbol{\psi}\left(\mathbf{r}\right)=\left(\begin{array}{l}
\frac{1}{\sqrt{3}}+\frac{1}{\sqrt{2}}\left(\psi_{1}^{\left(0\right)}+\frac{1}{\sqrt{3}}\psi_{2}^{\left(0\right)}\right)\\
\frac{1}{\sqrt{3}}+\frac{1}{\sqrt{2}}\left(-\psi_{1}^{\left(0\right)}+\frac{1}{\sqrt{3}}\psi_{2}^{\left(0\right)}\right)\\
\frac{1}{\sqrt{3}}\left(1-\sqrt{2}\psi_{2}^{\left(0\right)}\right)
\end{array}\right)
\end{eqnarray*}
with
\begin{eqnarray*}
\psi_{1}^{\left(0\right)} & =\frac{\lambda_{s}}{R^{2}}\frac{1}{\sqrt{6}}\left(\tilde{\xi}_{x}^{2}-\tilde{\xi}_{y}^{2}\right)r^{2},\\
\psi_{2}^{\left(0\right)} & =\frac{\lambda_{s}}{R^{2}}\frac{1}{3\sqrt{2}}\times\left[\left(\tilde{\xi}_{x}^{2}-\tilde{\xi}_{z}^{2}\right)+\left(\tilde{\xi}_{y}^{2}-\tilde{\xi}_{z}^{2}\right)\right]r^{2}
\end{eqnarray*}
where $\tilde{\mathbf{\xi}}\equiv\mathbf{\xi}/\xi=\left(\sin\theta\cos\varphi,\sin\theta\sin\varphi,\cos\theta\right)$
gives the direction of $\mathbf{\xi}$ within the nanomagnet.

More explicitly, we have
\begin{eqnarray}
\mathbf{m}\left(\xi,\theta,\varphi\right) & \cong\mathbf{m}_{0}+\frac{\lambda_{s}}{3\sqrt{12}}\left(\begin{array}{l}
\sin^{2}\theta\left(1+3\cos2\varphi\right)-2\cos^{2}\theta\\
\sin^{2}\theta\left(1-3\cos2\varphi\right)-2\cos^{2}\theta\\
-2\left(\sin^{2}\theta-2\cos^{2}\theta\right)
\end{array}\right)\xi^{2}.\label{eq:MagProfile-kc0}
\end{eqnarray}

This analytical result, a quadratic expansion in $\xi$, may be compared
to the numerical solution of the LLE (\ref{eq:LLE-MSP}). However,
such a comparison is not easy in practice and here is why.

The magnetization profile similar to Eq. (\ref{eq:MagProfile-kc0})
has been obtained, in the discrete approach, by solving the (damped)
LLE (\ref{eq:LLE-MSP}) for a spherical NM as defined earlier with
the Hamiltonian in Eq. (\ref{eq:Ham-MSP}). More precisely, we prepare
the NM by cutting a sphere in a simple-cubic $3D$ lattice of linear
size $N=N_{x}=N_{y}=N_{z}$, the outcome being a sphere-shaped ensemble
of $\mathcal{N}$ spins. Then, we set the physical parameters $J,K_{c},K_{s},h,$
etc, and run the Heun (or $4^{\mathrm{th}}$-order Runge-Kutta) routine
to solve Eq. (\ref{eq:LLE-MSP}), until the equilibrium state is reached.
The result is a spin configuration similar to that shown in Fig. \ref{fig:nsa-structure}.
For each such a spin configuration, we collect the spatial profile
of the net magnetic moment $\mathbf{m}$ as we go from the center
to the border of the NM, in a given direction. Now, because of the
discreteness of the underlying lattice (inside a spherical NM cut
out of a simple-cubic lattice), the raw profile, or the components
of $\mathbf{m}$ in Eq. (\ref{eq:Macrospin}) as a function of the
lattice site $\mathbf{r}_{i}$, yields rather jagged plots. In order
to smooth out the data, we may average over the direction $\left(\theta,\varphi\right)$
of $\mathbf{\xi}$ and consider only the radial profile of $\mathbf{m}$,
\emph{i.e.} $\mathbf{m}\left(\xi\right)$. This is given by

\begin{equation}
\left\langle \mathbf{m}\right\rangle _{\Omega}=\oint_{\partial V}\frac{d\Omega}{4\pi}\:\mathbf{m}\left(\boldsymbol{\xi}\right)=\mathbf{m}_{0}\times\left[1-\frac{1}{2}\oint_{\partial V}\frac{d\Omega}{4\pi}\|\boldsymbol{\psi}\left(\mathbf{r}\right)\|^{2}\right].\label{eq:OmegaAve_def}
\end{equation}

We see that upon averaging over the direction $\left(\theta,\varphi\right)$,
the linear contribution in $\boldsymbol{\psi}$ vanishes. This can
also be checked by performing the same average in Eq. (\ref{eq:psi-Coeffs0})
which, in turn, amounts to checking that the average of the GF (\ref{eq:GFn-vs})
over the direction of $\mathbf{\xi}$ vanishes. This result is consistent
with and justifies the condition (\ref{eq:Intpsizero}).

Then, the integration over $\Omega$ yields
\[
\oint_{\partial V}\frac{d\Omega}{4\pi}\left(\tilde{\xi}_{\alpha}^{2}-\tilde{\xi}_{\beta}^{2}\right)^{2}=\frac{4}{15},\quad\oint_{\partial V}\frac{d\Omega}{4\pi}\left(\tilde{\xi}_{x}^{2}-\tilde{\xi}_{z}^{2}\right)\left(\tilde{\xi}_{y}^{2}-\tilde{\xi}_{z}^{2}\right)=\frac{2}{15}
\]
and upon using $1=\left(m_{0,z}^{2}+m_{0,x}^{2}+m_{0,y}^{2}\right)^{2}$,
we obtain
\begin{eqnarray}
\oint_{\partial V}\frac{d\Omega}{4\pi}\|\boldsymbol{\psi}\left(\mathbf{r}\right)\|^{2} & =\frac{2}{15}\lambda_{s}^{2}\left(1-\sum_{\alpha}m_{0,\alpha}^{4}\right)\xi^{4}.\label{eq:OmegaAve-psi}
\end{eqnarray}

This finally leads to the solid-angle average of the magnetization
profile
\begin{eqnarray}
\left\langle \mathbf{m}\right\rangle _{\Omega}\left(\xi\right) & \simeq\mathbf{m}_{0}\left[1-\frac{\lambda_{s}^{2}}{15}\left(1-\sum_{\alpha}m_{0,\alpha}^{4}\right)\xi^{4}\right].\label{eq:OmegaAve_result}
\end{eqnarray}

Note that upon averaging over the direction $\left(\theta,\varphi\right)$,
the quadratic contribution in Eq. (\ref{eq:MagProfile-kc0}) vanishes
and only the quartic contribution remains in the expansion (\ref{eq:SecondOrderApproximation}).
The experimental techniques at our disposal today are not precise
enough to allow for a probe of the magnetization profile in a given
direction $\left(\theta,\varphi\right)$ within the nanomagnet. In
addition, even if this were possible, the prototypical nanomagnet
samples are assemblies with distributed nanomagnets and, as such only
an average over the whole assembly can be accessed by measurements.
This implies that if we were able to probe the magnetization profile,
we should most likely observe the quartic behavior given by Eq. (\ref{eq:OmegaAve_result}).

For later reference, we introduce the coefficient (of $\xi^{4}$)
\begin{equation}
\mathcal{C}_{\beta}\equiv m_{0,\beta}\frac{\lambda_{s}^{2}}{15}\left(1-\sum_{\alpha}m_{0,\alpha}^{4}\right),\quad\beta=x,y,z.\label{eq:MagProf_ave_coeff}
\end{equation}

In order to further smooth out the lattice-induced jaggedness of the
numerical data, we may also average over the magnitude of $\xi$ taken
within slices (or ring bands) perpendicular to the radial direction.
For this, we adopt an onion structure for the NM and plot the net
magnetic moment $\left\langle \mathbf{m}\right\rangle _{\Omega}$
as a function of the points $\xi_{n},i=1\ldots M$, each of which
being the center of a ring band. Doing so, leads to the discrete expression

\begin{eqnarray*}
\left\langle \mathbf{m}\right\rangle _{\Omega}^{n}\left(\xi_{n}\right) & \simeq\mathbf{m}_{0}-\mathbf{m}_{0}\frac{\lambda_{s}^{2}}{15}\left(1-\sum_{\alpha}m_{0,\alpha}^{4}\right)\xi_{n}^{4}
\end{eqnarray*}
or component-wise
\begin{eqnarray}
\left\langle m_{\beta}\right\rangle _{\Omega}^{n}\left(\xi_{n}\right) & \simeq m_{0,\beta}-\mathcal{C}_{\beta}\xi_{n}^{4}.\label{eq:MagProf_ave}
\end{eqnarray}

\begin{figure*}[h!]
\begin{centering}
\includegraphics[width=7.5cm]{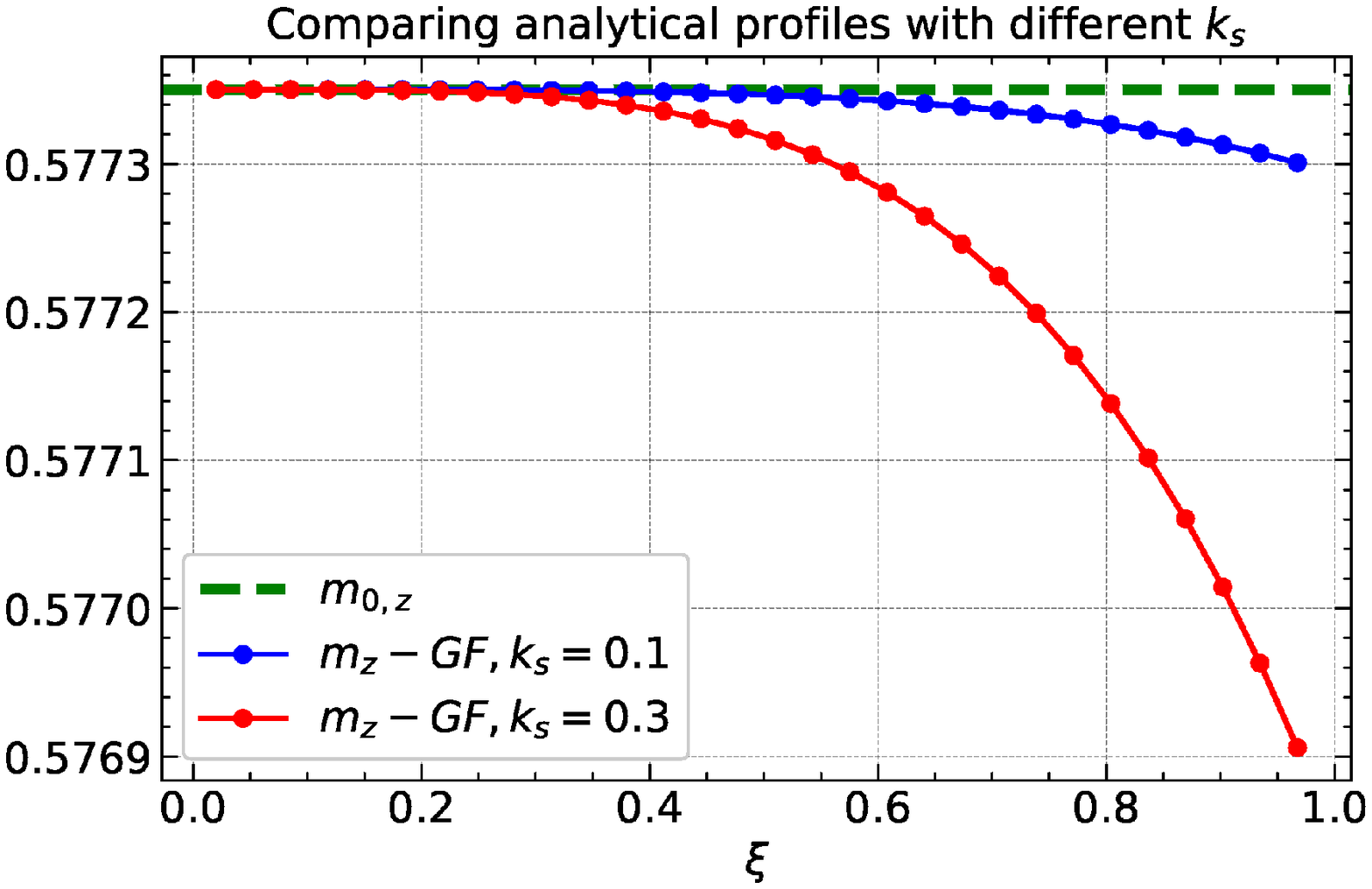} \includegraphics[width=7.5cm]{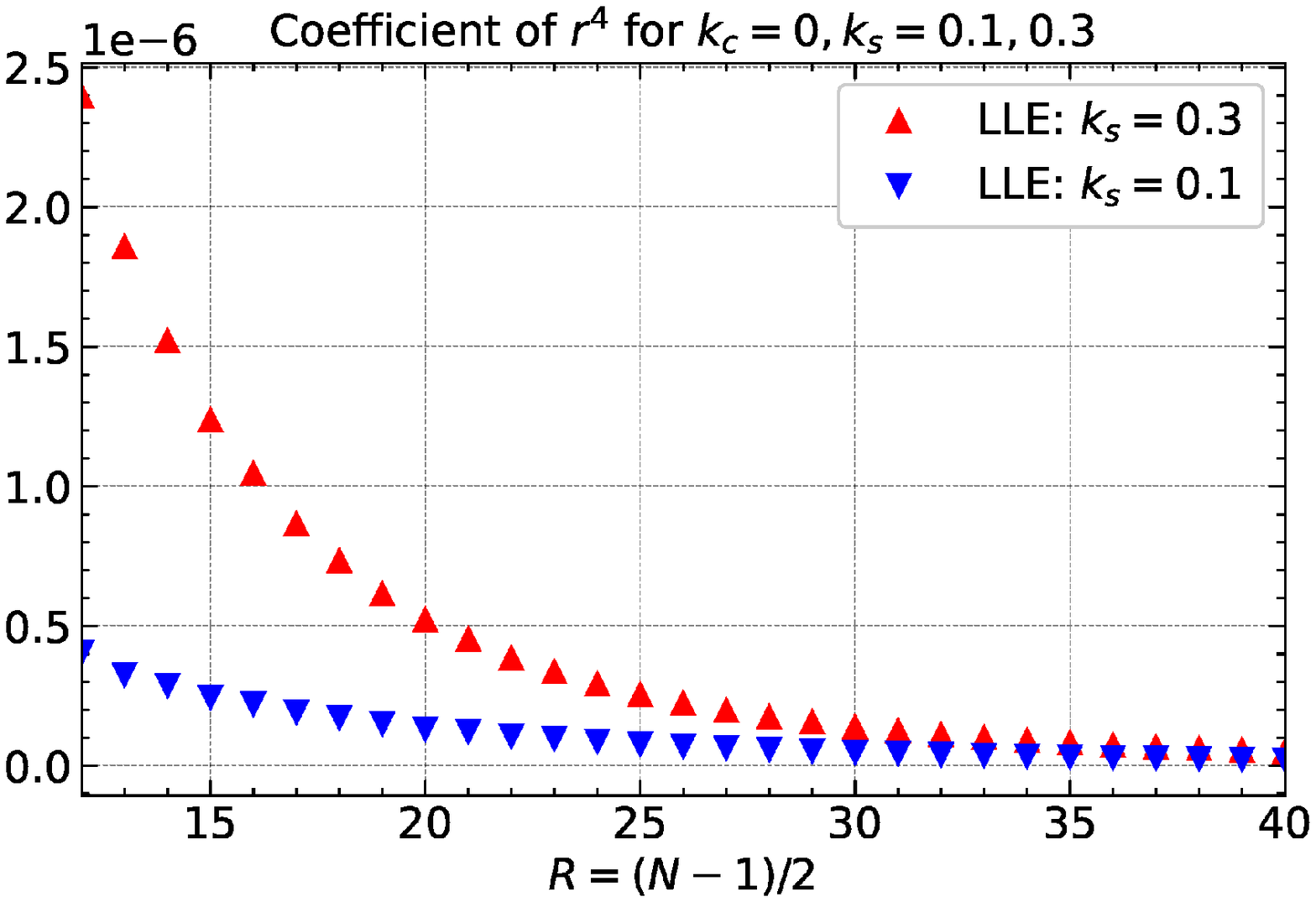}
\par\end{centering}
\caption{\label{fig:Coeffz} (Left) Magnetization profile $m_{z}$ against
$\xi$ given by Eq. (\ref{eq:MagProf_ave}) and (right) coefficient
of $r^{4}$ as a function of the NM radius $R=\left(N-1\right)/2$,
with $N$ being the linear size. This is $\mathcal{C}_{z}$ defined
in Eq. (\ref{eq:MagProf_ave_coeff}) divided by $R^{4}$. These results
are for a spherical NM with $k_{c}=0,k_{s}=0.1,0.3$. {[}No magnetic
field{]}.}
\end{figure*}

\begin{figure*}[h!]
\begin{centering}
\includegraphics[scale=0.6]{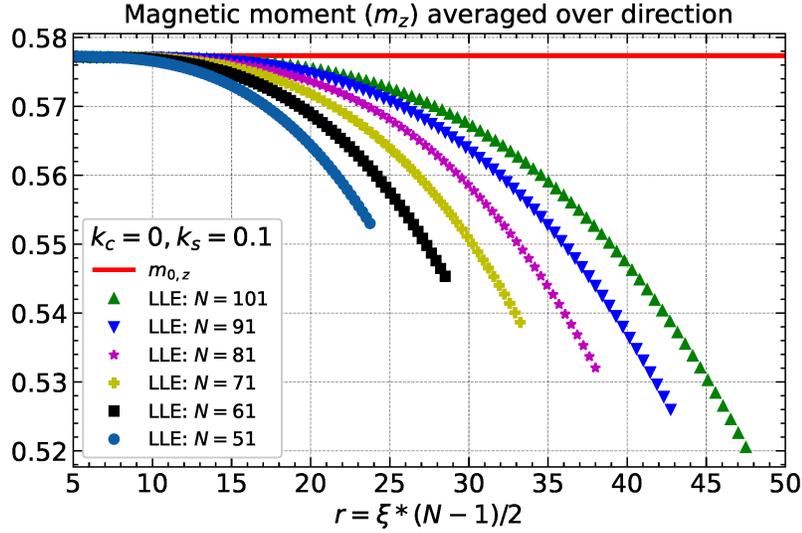}
\par\end{centering}
\caption{\label{fig:mz-GFvsLLE} Spatial profile of the $z$ component of the
NM net magnetic moment, averaged over the direction $\Omega$, as
a function of the radial distance $r=\left(N-1\right)/2\times\xi$,
with $0\protect\leq\xi\protect\leq1$, for $k_{c}=0,k_{s}=0.1$. The
continuous red line is the net magnetic moment component $m_{0,z}=1/\sqrt{3}$.
On the scale used here, the GF function curve given by Eq. (\ref{eq:MagProf_ave})
coincides with the asymptotic straight line $m_{0,z}=1/\sqrt{3}$.
{[}No magnetic field{]}.}
\end{figure*}

In the numerical calculations (numerical solution of the LLE), from
each spin configuration obtained for a set of physical parameters
and a given linear size $N$, we infer the average $\left\langle \mathbf{m}\right\rangle _{\Omega}^{n}\left(r_{n}=R\xi_{n}\right)$.
The latter is fit to $a_{\beta}-\mathcal{C}_{\beta}r^{4}$, for $\beta=x,y,z$,
to obtain the coefficient $\mathcal{C}_{\beta}$. The results are
shown in Fig. \ref{fig:Coeffz} (right). It is clearly seen that a
larger $k_{s}$ corresponds to a larger coefficient and thereby to
stronger spin misalignments or deviations from $\mathbf{m}_{0}$.
In addition, as the radius of the NM increases (we only show part
of the data that have been obtained for $N=25,26,\ldots,111$ or $R=12,13,\ldots,55$),
the coefficients for different values of $k_{s}$ tend to zero. Indeed,
as the size increases, the ratio of the number of surface spins to
the total number decreases to zero. This translates into negligible
surface effects and thereby to vanishing spin deviations. Indeed,
a fit of the curves in Fig. \ref{fig:Coeffz} (right) yields $\mathcal{C}_{\beta}\sim R^{-2}$,
leading to $\left\langle m_{\beta}\right\rangle _{\Omega}^{n}\left(r\right)\sim a_{\beta}-\left(b_{\beta}/R^{2}\right)r^{4}$,
where $b_{\beta}$ is a constant. This is illustrated in Fig. \ref{fig:mz-GFvsLLE}
where we compare the magnetic profile for different sizes to $m_{0,z}$,
the magnetic moment in the uniform state. Finally, it is worth noting,
by examining the vertical scale, that the deviation of the magnetic
moment from the net direction $\mathbf{m}_{0}$ is rather small but
it increases towards the NM boundary.

\subsection{\label{subsec:In-the-presence-of-Kc}In the presence of core anisotropy}

The more realistic situation with anisotropy in the core of the NM
($K_{{\rm c}}\neq0$), as well as on the surface ($K_{{\rm s}}\neq0$),
is more involved. Indeed, there is no GF solving the problem stated
in Eqs. (\ref{eq:GF_Helmholtz_general}, \ref{eq:g-BC}). However,
since the coefficients $\kappa_{\alpha}^{2}$ in Eq. (\ref{eq:HelmholtzCoefficients})
are small, owing to the fact that the core anisotropy and the applied
field are, in typical situations, small with respect to the exchange
coupling, we can use a perturbative approach. Indeed, we may write
\cite{kachkachi07j3m}

\begin{equation}
\psi_{\beta}\left(\boldsymbol{\xi}\right)\simeq\psi_{\beta}^{\left(0\right)}\left(\boldsymbol{\xi}\right)+\kappa_{\beta}^{2}\,\psi_{\beta}^{\left(1\right)}\left(\boldsymbol{\xi}\right).\label{eq:psibeta-expansion}
\end{equation}

Then, substituting in Eq. (\ref{eq:DecoupledHelmholtzEquation}),
using $\Delta_{\xi}\psi_{\beta}^{\left(0\right)}\left(\boldsymbol{\xi}\right)=0$
{[}see Section \ref{subsecKcZero}{]} and dropping the term in $\kappa_{\beta}^{4}$,
leads to

\[
\Delta_{\xi}\psi_{\beta}^{\left(1\right)}\left(\boldsymbol{\xi}\right)=\psi_{\beta}^{\left(0\right)}\left(\boldsymbol{\xi}\right).
\]

Next, we have
\[
\left.\frac{\mathrm{d}\psi_{\beta}}{\mathrm{d}\xi}\right|_{\xi=1}=\left.\frac{\mathrm{d}\psi_{\beta}^{\left(0\right)}}{\mathrm{d}\xi}\right|_{\xi=1}+\kappa_{\beta}^{2}\,\left.\frac{\mathrm{d}\psi_{\beta}^{\left(1\right)}}{\mathrm{d}\xi}\right|_{\xi=1}.
\]

Now, $\boldsymbol{\psi}^{\left(0\right)}$ is the major contribution to
$\boldsymbol{\psi}$ that stems from surface anisotropy and $\boldsymbol{\psi}^{\left(1\right)}$
appears only in the presence of core anisotropy and/or applied magnetic
field ($\kappa_{\beta}\neq0$), which tend to reduce the spin misalignments.
We may then consider that $\boldsymbol{\psi}^{\left(0\right)}$ still
satisfies the boundary conditions (\ref{eq:NeumannBC}), thus leading
to
\[
\left.\frac{\mathrm{d}\psi_{\beta}^{\left(1\right)}}{\mathrm{d}\xi}\right|_{\xi=1}=0.
\]

Therefore, $\psi_{\beta}^{\left(1\right)}$ is a field that satisfies
Poisson's equation subjected to homogeneous Neumann boundary conditions,
namely

\begin{eqnarray}
\Delta_{\xi}\psi_{\beta}^{\left(1\right)}\left(\boldsymbol{\xi}\right) & =\psi_{\beta}^{\left(0\right)}\left(\boldsymbol{\xi}\right),\label{eq:psi1-pb-eqt}\\
\left.\frac{\mathrm{d}\psi_{\beta}^{\left(1\right)}}{\mathrm{d}\xi}\right|_{\xi=1} & =0.\label{eq:psi1-pb-bc}
\end{eqnarray}

The solution of this problem can only exist if $\int_{V}d\boldsymbol{\xi}\,\psi_{\beta}^{\left(0\right)}\left(\boldsymbol{\xi}\right)=0$.
It can be checked that this is indeed the case by using expressions
(\ref{eq:psi0-12-sol}). This is also compatible with the condition
(\ref{eq:Intpsizero}) that could be assumed to apply at all orders
of perturbation. In this case, there exists a GF, call it $\tilde{\mathcal{G}}\left(\boldsymbol{\xi}^{\prime},\boldsymbol{\xi}\right)$,
satisfying
\begin{equation}
\left\{ \begin{array}{lll}
\Delta_{\xi}\tilde{\mathcal{G}}\left(\boldsymbol{\xi}^{\prime},\boldsymbol{\xi}\right) & = & -4\pi\delta\left(\boldsymbol{\xi}-\boldsymbol{\xi}'\right),\\
\\
\left.\frac{\mathrm{d}\tilde{\mathcal{G}}\left(\boldsymbol{\xi}^{\prime},\boldsymbol{\xi}\right)}{\mathrm{d}\xi}\right|_{\xi=1} & = & -1.
\end{array}\right.\label{eq:gtilde}
\end{equation}

The solution of the problem then reads \cite{MorseFeschbach_mgh53}
\begin{eqnarray*}
\psi_{\beta}^{\left(1\right)}\left(\boldsymbol{\xi}\right) & =-\frac{1}{4\pi}\int_{V}d\boldsymbol{\xi}^{\prime}\,\tilde{\mathcal{G}}\left(\boldsymbol{\xi}^{\prime},\boldsymbol{\xi}\right)\psi_{\beta}^{\left(0\right)}\left(\boldsymbol{\xi}^{\prime}\right)-\frac{1}{4\pi}\oint_{\partial V}d^{2}n^{\prime}\,\psi_{\beta}^{\left(1\right)}\left(\mathbf{n}^{\prime}\right)\left.\frac{\mathrm{d}\tilde{\mathcal{G}}\left(\boldsymbol{\xi}^{\prime},\boldsymbol{\xi}\right)}{\mathrm{d}\xi'}\right|_{\xi^{\prime}=1}\\
 & =-\frac{1}{4\pi}\int_{V}d\boldsymbol{\xi}^{\prime}\,\tilde{\mathcal{G}}\left(\boldsymbol{\xi}^{\prime},\boldsymbol{\xi}\right)\psi_{\beta}^{\left(0\right)}\left(\boldsymbol{\xi}^{\prime}\right)+ D_{\beta},
\end{eqnarray*}
where $D_{\beta}$ is a constant. In fact, we see that $\tilde{\mathcal{G}}\left(\boldsymbol{\xi},\boldsymbol{\xi}^{\prime}\right)$
is the solution of the same problem as $\mathcal{G}^{(0)}\left(\boldsymbol{\xi},\boldsymbol{\xi}^{\prime}\right)$
and, as such, we may simply take $\tilde{\mathcal{G}}\left(\boldsymbol{\xi},\boldsymbol{\xi}^{\prime}\right)=\mathcal{G}^{(0)}\left(\boldsymbol{\xi},\boldsymbol{\xi}^{\prime}\right)$.
In addition, the constant $D_{\beta}$ can be determined
by assuming that the spin mis-alignment vanishes at the center of
the NM, \emph{i.e.} $\psi_{\beta}^{\left(1\right)}\left(\boldsymbol{0}\right)=0$.
This yields, using Eq. (\ref{eq:GF0-rrp}), $D_{\beta}=\frac{1}{4\pi}\int_{V}d\boldsymbol{\xi}\,\mathcal{G}^{(0)}\left(0,\boldsymbol{\xi}\right)\psi_{\beta}^{\left(0\right)}\left(\boldsymbol{\xi}\right)=0$.

Finally, we obtain the solution
\begin{eqnarray}
\psi_{\beta}^{\left(1\right)}\left(\boldsymbol{\xi}\right) & =-\frac{1}{4\pi}\int_{V}d\boldsymbol{\xi}^{\prime}\,\mathcal{G}^{(0)}\left(\boldsymbol{\xi},\boldsymbol{\xi}^{\prime}\right)\psi_{\beta}^{\left(0\right)}\left(\boldsymbol{\xi}^{\prime}\right).\label{eq:psi1-sol}
\end{eqnarray}

Note that this result can also be obtained by proceeding through an
expansion of the GF $\mathcal{G}_{\beta}\left(\boldsymbol{\xi},\mathbf{n}^{\prime}\right)$
that appears in Eq. (\ref{eq:GF_Helmholtz_general}), instead of the
expansion in Eq. (\ref{eq:psibeta-expansion}). This is done in
\ref{sec:Expansion-of-GF-g}. Eq. (\ref{eq:psi1-sol}), which derives
from Eq. (\ref{eq:psi1-pb-eqt}), suggests that $\psi_{\beta}^{\left(0\right)}$
acts as a source for the field $\psi_{\beta}^{\left(1\right)}$.

Let us now discuss the explicit calculation of the components $\psi_{\beta}^{\left(1\right)}\left(\boldsymbol{\xi}\right)$
of the spin deviation. Note that in Eq. (\ref{eq:psi1-sol}), we have
an integral over the volume and thereby none of the arguments of $\mathcal{G}^{(0)}\left(\boldsymbol{\xi},\boldsymbol{\xi}^{\prime}\right)$
is fixed on the surface. As a consequence, we have to use the exact
expression (\ref{eq:GF0-rrp}), instead of the expansion (\ref{eq:GF0-Expansion}).
Unfortunately, it is then difficult to obtain a closed analytical
result for the integral in Eq. (\ref{eq:psi1-sol}). On the other
hand, if we use instead the representation (\ref{eq:psi1-sol}) in
terms of the GF $\mathcal{G}^{(1)}$ in Eq. (\ref{eq:G1Res}), we
again encounter an integral over the volume of the product of two
$\mathcal{G}^{(0)}$, one of which has both arguments inside $\partial V$.
Consequently, we can provide analytical (approximate) expressions
for $\psi_{\beta}^{\left(1\right)}\left(\boldsymbol{\xi}\right)$
only for $\boldsymbol{\xi}$ on the boundary $\partial V$. This should
yield the largest contribution from $\psi_{\beta}^{\left(1\right)}\left(\boldsymbol{\xi}\right)$,
as one obtains for $\psi_{\beta}^{\left(0\right)}$ in Eq. (\ref{eq:psi0-LargeDev}),
see below. For arbitrary $\boldsymbol{\xi}$, with $0\leq\xi\leq1$,
we must resort to numerical integration.

For $\boldsymbol{\xi}$ on the boundary $\partial V$, \emph{i.e.
}$\boldsymbol{\xi}=\mathbf{n}$, we use Eqs. (\ref{eq:GF0-Expansion})
and (\ref{eq:psi0-LargeDev}) to derive the following expressions
for the components of $\boldsymbol{\psi}^{\left(1\right)}$ on the sphere:

\begin{eqnarray}
\psi_{1}^{\left(1\right)}\left(\mathbf{n},\mathbf{m}_{0}\right) & \simeq-\frac{\lambda_{s}}{14}\frac{m_{0,x}m_{0,y}}{\sqrt{1-m_{0,z}^{2}}}\left(n_{y}^{2}-n_{x}^{2}\right),\label{eq:psi1-LargeDev}\\
\psi_{2}^{\left(1\right)}\left(\mathbf{n},\mathbf{m}_{0}\right) & \simeq-\frac{\lambda_{s}}{14}\frac{m_{0,z}}{\sqrt{1-m_{0,z}^{2}}}\left[\left(n_{x}^{2}-n_{z}^{2}\right)m_{0,x}^{2}+\left(n_{y}^{2}-n_{z}^{2}\right)m_{0,y}^{2}\right].\nonumber
\end{eqnarray}

The components of the largest spin deviation represented by the (total)
vector $\boldsymbol{\psi}$, within a spherical NM with equilibrium magnetic
moment $\mathbf{m}_{0}$, are obtained by substituting (\ref{eq:psi0-LargeDev})
and (\ref{eq:psi1-LargeDev}) into Eq. (\ref{eq:psibeta-expansion}).
This yields

\begin{eqnarray}
\psi_{1}\left(\mathbf{n},\mathbf{m}_{0}\right) & \simeq\lambda_{s}\left(1-\frac{\kappa_{1}^{2}}{14}\right)\frac{m_{0,x}m_{0,y}}{\sqrt{1-m_{0,z}^{2}}}\left(n_{x}^{2}-n_{y}^{2}\right),\label{eq:psi-LargestDev}\\
\psi_{2}\left(\mathbf{n},\mathbf{m}_{0}\right) & \simeq\lambda_{s}\left(1-\frac{\kappa_{2}^{2}}{14}\right)\frac{m_{0,z}}{\sqrt{1-m_{0,z}^{2}}}\left[\left(n_{x}^{2}-n_{z}^{2}\right)m_{0,x}^{2}+\left(n_{y}^{2}-n_{z}^{2}\right)m_{0,y}^{2}\right].\nonumber
\end{eqnarray}

Note that because of the factor $1-\kappa_{\alpha}^{2}/14$, these
expressions are valid for $\kappa_{\alpha}\leq\sqrt{14}\simeq3.74$.
However, since they have been derived using an expansion in $\kappa_{\alpha}$,
these expressions are actually valid for a much smaller $\kappa_{\alpha}$
and the previous condition adds no new constraint.

\begin{figure*}[h!]
\begin{centering}
\includegraphics[scale=0.3]{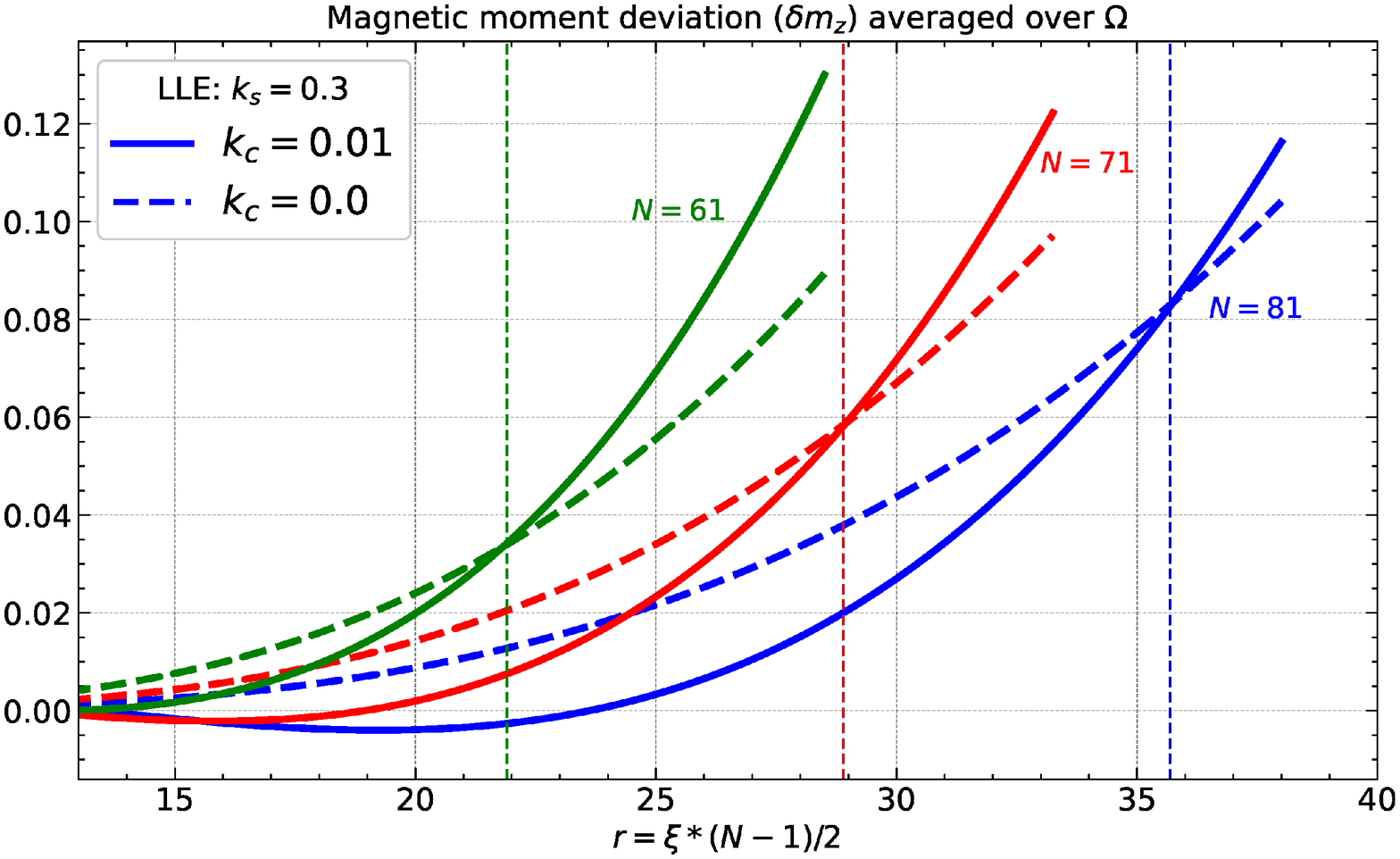} \includegraphics[scale=0.3]{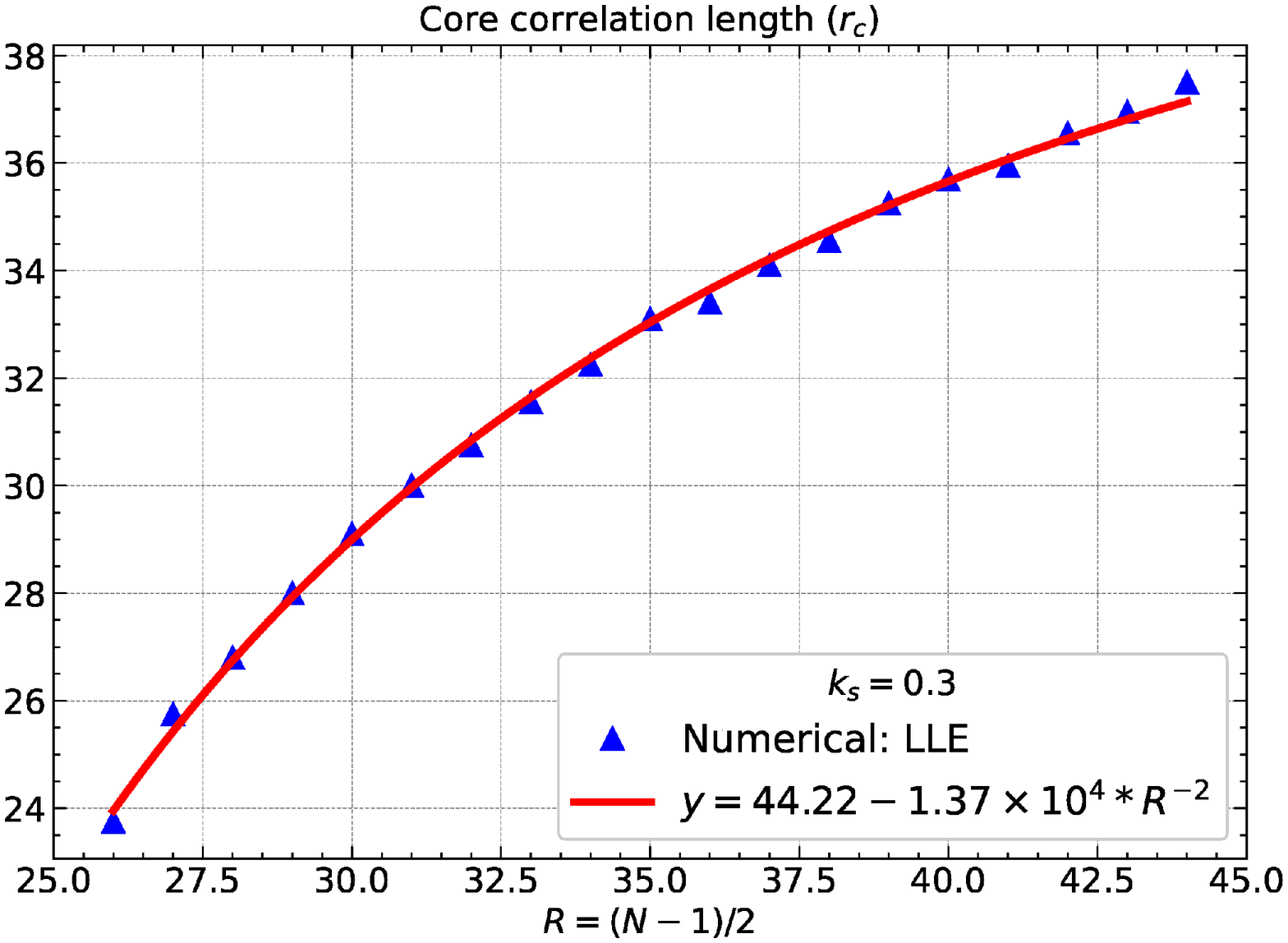}
\par\end{centering}
\caption{\label{fig:mz-LLE-kc0kc001-ks03} (Left) Deviation of the net magnetic
moment $\delta m_{z}\equiv m_{z}\left(r\right)-m_{z}\left(0\right)$,
averaged over the direction $\Omega$, as a function of the radial
distance $r=\left(N-1\right)/2\times\xi$, for $k_{s}=0.3$ and $k_{c}=0.01$
(full lines) and $k_{c}=0$ (dashed lines). Note that, for the reasons
explained in the text, the maximum value of $\xi$ is $0.95$ and
this is why the curves do not reach the last points at $r=30,35,40$.
(Right) The core correlation length over which the core anisotropy
dominates (see text). {[}No magnetic field{]}.}
\end{figure*}

We can also numerically compute the integral in (\ref{eq:psi1-sol})
and then average over the solid angle. However, although this procedure
is quite affordable to today's computers using optimized algorithms,
it still remains rather costly with regard to the CPU resources, especially
when several curves are needed for comparison. Here, we resort to
the numerical solution of the LLE system, as done in the case of $k_{c}=0$,
which allows for the full procedure in an easier manner. Accordingly,
in Fig. \ref{fig:mz-LLE-kc0kc001-ks03} we plot the deviation of the
$z$ component of the net magnetic moment, $\delta m_{z}\equiv m_{z}\left(r\right)-m_{z}\left(0\right)$,
averaged over the direction $\Omega$, as a function of the radial
distance $r=\left(N-1\right)/2\times\xi$, for $k_{s}=0.3$ and $k_{c}=0.01$
(full lines) and $k_{c}=0$ (dashed lines). We recall that the uniaxial
anisotropy here is taken along the $z$ axis. If it is taken along
the cube diagonal, the deviations will be much smaller. Indeed, we
note that in the presence of anisotropy in the core with an easy axis
in the $z$ direction, the state $\mathbf{m}_{0}$ is no longer along
the cube diagonal; it is tilted towards the $z$ axis by an angle
that depends on the relative strength of the core anisotropy ($k_{c}$).
This results in a competition between the core and surface anisotropies.

The results in Fig. \ref{fig:mz-LLE-kc0kc001-ks03} show that, both
with and without core anisotropy, the overall spin deviations are
reduced when the NM size increases. In addition, here we see that
for a given size $N$ of the NM, the curves with and without core
anisotropy (same color) intersect at a given distance $r_{c}$ from
the center of the NM. As discussed earlier, the spin misalignments
induced by the surface anisotropy tend to propagate from the boundary
to the center of the NM, while the effect of the core (uniaxial) anisotropy
is to align the spins parallel to each other and thus to push the
spin misalignments out to the border. The competition between these
two effects results in a critical radius $r_{c}$, or core correlation
length (indicated by the dashed vertical lines), over which the core
anisotropy dominates, thereby rendering a weaker spin deviation. This
is illustrated by the fact that, for $r\leq r_{c}$, the continuous
curves ($k_{c}\neq0$) are below the dashed ones ($k_{c}=0$). Furthermore,
in the plot on the right, we see that the distance $r_{c}$ increases
with the radius $R=\left(N-1\right)/2$ of the NM ; it behaves as
$a-b/R^{2}$ with $b>0$ (see the fitting curve in red). So, as $R$
increases the surface relative contribution decreases and the core
anisotropy then dominates and pushes the spin noncollinearities farther
out towards the NM border. As a consequence, the surface contribution
to the overall anisotropy of the NM scales with the surface ($\sim R^{2}$),
as was discussed in Ref. \cite{garkac03prl}.

\section{\label{sec:Summary-Conclusions}Summary, Conclusions, and Outlook}

We have built a formalism for solving the Helmholtz equation, with
inhomogeneous Neumann boundary conditions, satisfied by the spin deviation
vector induced by surface anisotropy in a nanomagnet, using the technique
of Green's functions in the continuum limit. The nanomagnet has been
modeled as a spherical crystallite of $\mathcal{N}$ atomic magnetic
moments and whose energy comprises the exchange interaction, the Zeeman
contribution and the anisotropy energy that discriminates between
spins in the core, attributed a uniaxial anisotropy, and spins at
the surface whose anisotropy is given by N\'eel's model. We have also
provided the numerical solution of a system of coupled Landau-Lifshitz
equations written for the atomic magnetic moments and compared the
results to those of the analytical approach. We have computed the
solid-angle averaged components of the nanomagnet's net moment as
a function of the distance to the NM center in the radial direction,
both in the absence and presence of anisotropy in the core. In the
former case, we have provided good approximate analytical expressions
for the spin deviation at an arbitrary position within the nanomagnet.
In the latter case, however, the solution is only given numerically,
either through a volume integral within the Green's function approach,
or numerically by solving the Landau-Lifshitz equations. Nonetheless,
an analytical solution for this case has been given on the boundary
of the NM, which represents the largest spin deviation. Both the numerical
and (semi-)analytical results show that the spin deviations induced
by surface anisotropy are stronger with larger surface anisotropy
constant and/or smaller sizes.

As discussed in the introduction, the small-angle neutron scattering
technique should provide us with a relatively precise probe of a signature
of spin deviations in nanomagnets. However, with real samples, we
are faced with various distributions (size, shape and anisotropy)
and collective effects due to inter-particle interactions which may
lead to a smearing out of the surface effects and the entailed sought-for
spin misalignments. As a first step, we may consider doing measurements
on an array of well separated platelets (or thin cylinders), thus
avoiding strong inter-particle interactions while ensuring enhanced
surface contributions to the overall anisotropy. In parallel to these
investigations, further theoretical endeavor is required in order
to take account of the inter-particle interactions together with other
forms of anisotropy that might stem from different shapes and internal
structures of the nanomagnets (\emph{e.g.} platelets). In this context,
the present Green's function methodology may form the basis for computing
the magnetic small angle neutron scattering cross-section of nanomagnets
according to their magnetic materials parameters.

\newpage{}

\newpage{}

\newpage\appendix

\section{\label{sec:Expansion-of-GF-g}Expansion of the Green's function in the presence of core anisotropy}

Similarly to the expansion of $\boldsymbol{\psi}$ in Eq. (\ref{eq:psibeta-expansion}),
we may write the GF $\mathcal{G}_{\beta}$ that appears in Eq. (\ref{eq:GF_Helmholtz_general})
as follows \cite{kachkachi07j3m}
\begin{equation}
\mathcal{G}_{\alpha}\left(\mathbf{m}_{0},\boldsymbol{\xi},\boldsymbol{\xi}^{\prime}\right)=\mathcal{G}^{(0)}\left(\boldsymbol{\xi},\boldsymbol{\xi}^{\prime}\right)+\kappa_{\alpha}^{2}\,\left(\mathbf{m}_{0},\mathbf{h}\right)\mathcal{G}^{(1)}\left(\boldsymbol{\xi},\boldsymbol{\xi}^{\prime}\right)+\ldots.\label{eq:ExpandG0}
\end{equation}

It is then easy to see, upon using Eq. (\ref{eq:0thOrderGF}), that
the correction term $\mathcal{G}^{(1)}\left(\boldsymbol{\xi},\boldsymbol{\xi}'\right)$
satisfies the following equation (upon dropping terms in $\kappa_{\alpha}^{4}$)
\begin{equation}
\Delta\mathcal{G}^{(1)}\left(\boldsymbol{\xi},\boldsymbol{\xi}^{\prime}\right)\simeq\mathcal{G}^{(0)}\left(\boldsymbol{\xi},\boldsymbol{\xi}^{\prime}\right)\label{eq:RelationGF0GF1}
\end{equation}
and that its solution can be written as a convolution
\begin{equation}
\mathcal{G}^{(1)}\left(\boldsymbol{\xi},\boldsymbol{\xi}^{\prime}\right)=-\frac{1}{4\pi}\int_{V}\mathcal{G}^{(0)}\left(\boldsymbol{\xi}^{\prime\prime},\boldsymbol{\xi}^{\prime}\right)\mathcal{G}^{(0)}\left(\boldsymbol{\xi},\boldsymbol{\xi}^{\prime\prime}\right)\;d^{3}\xi^{\prime\prime}\label{eq:G1Res}
\end{equation}
with the boundary condition [using Eq. (\ref{eq:g0-BC})]
\begin{eqnarray}
\left.\frac{\mathrm{d}\mathcal{G}^{(1)}}{\mathrm{d}\xi}\right|_{\xi=1} & =\frac{1}{4\pi}\int_{V}\mathcal{G}^{(0)}\left(\boldsymbol{\xi}^{\prime\prime},\boldsymbol{\xi}^{\prime}\right)\;\mathrm{d}^{3}\xi^{\prime\prime}.\label{eq:g1-BC}
\end{eqnarray}

Then, in Eq. (\ref{eq:GF_Helmholtz_general3}), we substitute the
expansions for $\psi_{\beta}\left(\boldsymbol{\xi}\right)$ and $\mathcal{G}_{\beta}\left(\boldsymbol{\xi},\mathbf{n}^{\prime}\right)$,
from Eqs. (\ref{eq:psibeta-expansion}) and (\ref{eq:ExpandG0}),
respectively, and identifying the terms of the same order in $\kappa_{\beta}$,
we obtain the following two equations
\begin{eqnarray*}
\psi_{\beta}^{\left(0\right)}\left(\boldsymbol{\xi}\right) & =\frac{1}{4\pi}\oint_{\partial V}d^{2}n^{\prime}\left.\frac{\mathrm{d}\psi_{\beta}}{\mathrm{d}\xi}\right|_{\boldsymbol{\xi}=\mathbf{n}^{\prime}}\mathcal{G}^{(0)}\left(\boldsymbol{\xi},\mathbf{n}^{\prime}\right)\\
\psi_{\beta}^{\left(1\right)}\left(\boldsymbol{\xi}\right) & =\frac{1}{4\pi}\oint_{\partial V}d^{2}n^{\prime}\left.\frac{\mathrm{d}\psi_{\beta}}{\mathrm{d}\xi}\right|_{\boldsymbol{\xi}=\mathbf{n}^{\prime}}\mathcal{G}^{(1)}\left(\boldsymbol{\xi},\mathbf{n}^{\prime}\right).
\end{eqnarray*}

Next, using (\ref{eq:psi1-pb-bc}) we recover Eq. (\ref{eq:psi-Coeffs0})
for the component $\psi_{\beta}^{\left(0\right)}$ together with the
following equation for $\psi_{\beta}^{\left(1\right)}$:

\begin{eqnarray}
\psi_{\beta}^{\left(1\right)}\left(\boldsymbol{\xi}\right) & =\frac{1}{4\pi}\oint_{\partial V}d^{2}n^{\prime}\Sigma_{\beta}\left(\mathbf{m}_{0},\mathbf{n}^{\prime}\right)\mathcal{G}^{(1)}\left(\boldsymbol{\xi},\mathbf{n}^{\prime}\right).\label{eq:psi1-in-g1}
\end{eqnarray}

Then, replacing $\mathcal{G}^{(1)}\left(\boldsymbol{\xi},\mathbf{n}^{\prime}\right)$
by its expression in Eq. (\ref{eq:G1Res}) leads to
\begin{equation*}
\psi_{\beta}^{\left(1\right)}\left(\boldsymbol{\xi}\right) =-\frac{1}{4\pi}\int_{V}d^{3}\xi^{\prime\prime}\,\mathcal{G}^{(0)}\left(\boldsymbol{\xi},\boldsymbol{\xi}^{\prime\prime}\right)
\left[\frac{1}{4\pi}\oint_{\partial V}d^{2}n^{\prime}\,\Sigma_{\beta}\left(\mathbf{m}_{0},\mathbf{n}^{\prime}\right)\mathcal{G}^{(0)}\left(\boldsymbol{\xi}^{\prime\prime},\mathbf{n}^{\prime}\right)\right].
\end{equation*}

Here, we recognize the term between brackets as $\psi_{\beta}^{\left(0\right)}$,
according to Eq. (\ref{eq:psi-Coeffs0}), thus recovering (up to a
constant) the result obtained in Eq. (\ref{eq:psi1-sol}). Note that
the main difference between the two representations, is that (\ref{eq:psi1-sol})
is an integral over the volume of the NM whereas (\ref{eq:psi1-in-g1})
is an integral over its surface.

\newpage{}
\ack M. Adams and A. Michels thank the National Research Fund of Luxembourg for financial support (AFR Grant No. 15639149).
H. Kachkachi thanks F. Vernay for reading the manuscript and suggesting improvements.

\end{document}